\def\year{2022}\relax
\documentclass[letterpaper]{article} 
\usepackage{aaai22}  
\usepackage{times}  
\usepackage{helvet}  
\usepackage{courier}  
\usepackage[hyphens]{url}  
\usepackage{graphicx} 
\usepackage{natbib}  
\usepackage{caption} 
\DeclareCaptionStyle{ruled}{labelfont=normalfont,labelsep=colon,strut=off} 
\usepackage{amsthm}
\newtheorem{theorem}{Theorem}
\newtheorem{prop}{Proposition}

\usepackage{aaai22}
\usepackage{times}
\usepackage{helvet}
\usepackage{courier}
\usepackage{graphicx}
\usepackage{float}
\usepackage[switch]{lineno}
\usepackage{multirow}
\usepackage{adjustbox}
\usepackage{amsmath}
\usepackage{amssymb}
\usepackage{xcolor}
\usepackage{algorithmic}
\usepackage[ruled,vlined]{algorithm2e}
\usepackage{subcaption}
\usepackage{kotex} 
\usepackage{comment}

\definecolor{ao}{rgb}{0.0, 0.5, 0.0}

\newcommand{\argmax}{\text{ArgMax}}
\newcommand{\leakyrelu}[1]{\text{LReLU}}

\newcommand{\floor}[1]{\left\lfloor #1 \right\rfloor}

\newcommand{\uniform}{\textbf{Unif}}

\newcommand{\test}{\mathbf{Test}}

\newcommand{\iggycmt}[1]{{\color{blue}{[Hyunsung] #1}}}
\usepackage{appendix}

\begin{document}

\author{
    Honguk Woo\textsuperscript{\rm 1}\thanks{Honguk Woo is the corresponding author.},
    Hyunsung Lee\textsuperscript{\rm 2} \equalcontrib,
    Sangwoo Cho\textsuperscript{\rm 1} \equalcontrib
}

\affiliations{
    \textsuperscript{\rm 1}Department of Computer Science and Engineering, Sungkyunkwan University\\ 
    \textsuperscript{\rm 2}Kakao Corporation\\
    hwoo@skku.edu, iggy.ll@kakaocorp.com, jsw7460@skku.edu \\
}

\title{An Efficient Combinatorial Optimization Model Using Learning-to-Rank Distillation}

\maketitle

\begin{abstract}
Recently, deep reinforcement learning (RL) has proven its feasibility in solving combinatorial optimization problems (COPs). The learning-to-rank techniques have been studied in the field of information retrieval. While several COPs can be formulated as the prioritization of input items, as is common in the information retrieval, it has not been fully explored how the learning-to-rank techniques can be incorporated into deep RL for COPs.  
In this paper, we present the learning-to-rank distillation-based COP framework, where a high-performance ranking policy obtained by RL for a COP can be distilled into a non-iterative, simple model, thereby achieving a low-latency COP solver. Specifically, we employ the approximated ranking distillation to render a score-based ranking model learnable via gradient descent. Furthermore, we use the efficient sequence sampling to improve the inference performance with a limited delay. 
With the framework, we demonstrate that a distilled model not only achieves comparable performance to its respective, high-performance RL, but also provides several times faster inferences. We evaluate the framework with several COPs such as priority-based task scheduling and multidimensional knapsack, demonstrating the benefits of the framework in terms of inference latency and performance.  
\end{abstract}

\section{Introduction} \label{sec:introduction}
In the field of computer science, it is considered challenging to tackle combinatorial optimization problems (COPs) that are computationally intractable. 
While numerous heuristic approaches have been studied to provide polynomial-time solutions, they often require in-depth knowledge on problem-specific features and customization upon the changes of problem conditions.  Furthermore, several heuristic approaches such as branching~\citep{chu1998genetic} and tabu-search~\cite{glover1989tabu} to solving COPs explore combinatorial search spaces extensively, and thus render themselves limited in large scale problems. 

Recently, deep learning techniques have proven their feasibility in addressing COPs, e.g., routing optimization~\cite{kool2018attention}, task scheduling~\cite{lee2020panda}, and knapsack problem~\cite{gupointerkp}. 
For deep learning-based COP approaches, it is challenging to build a training dataset with optimal labels, because many COPs are computationally infeasible to find exact solutions. Reinforcement learning (RL) is considered viable for such problems as neural architecture search~\cite{zoph2016neural}, device placement~\cite{mirhoseini2017device}, games~\cite{silver2017mastering}  where collecting supervised labels is expensive or infeasible.

As the RL action space of COPs can be intractably large (e.g., $100!$ possible solutions for ranking 100-items), it is undesirable to use a single probability distribution on the whole action space. Thus, a sequential structure, in which a probability distribution of an item to be selected next is iteratively calculated to represent a one-step action, becomes a feasible mechanism to establish RL-based COP solvers, as have been recently studied in~\cite{bello2016neural,vinyals2019grandmaster}. 
The sequential structure is effective to produce a permutation comparable to optimal solutions, but it often suffers from long inference time due to its iterative nature. Therefore, it is not suitable to apply these approaches to the field of mission critical applications with strict service level objectives and time constraints. For example, task placement in SoC devices necessitates fast inferences in a few milliseconds, but the inferences by a complex model with sequential processing often take a few seconds, so it is rarely feasible to employ deep learning-based task placement in SoC~\cite{ykman2006fast,shojaei2009parameterized}.

In this paper, we present \textbf{RLRD}, an RL-to-rank distillation framework to address COPs, which enables the low-latency inference in online system environments. To do so, we develop a novel ranking distillation method and focus on two COPs where each problem instance can be treated as establishing the optimal policy about ranking items or making priority orders. Specifically, we employ a differentiable relaxation scheme for sorting and ranking operations~\cite{blondel2020fast} to expedite direct optimization of ranking objectives. It is combined with a problem-specific objective to formulate a distillation loss that corrects the rankings of input items, thus enabling the robust distillation of the ranking policy from sequential RL to a non-iterative, score-based ranking model. 
Furthermore, we explore the efficient sampling technique with Gumbel trick~\cite{jang2016categorical, kool2019estimating} on the scores generated by the distilled model to expedite the generation of sequence samples and improve the model performance with an inference time limitation.    

Through experiments, we demonstrate that a distilled model by the framework achieves the low-latency inference while maintaining comparable performance to its teacher. For example, compared to the high-performance teacher model with sequential RL, the distilled model makes inferences up to 65 times and 3.8 times faster, respectively for the knapsack and task scheduling problems, while it shows only about 2.6\% and 1.0\% degradation in performance, respectively. The Gumbel trick-based sequence sampling improves the performance of distilled models (e.g., 2\% for the knapsack) efficiently with relatively small inference delay. 
The contributions of this paper are summarized as follows. 
\begin{itemize}
 \item We present the learning-based efficient COP framework RLRD that can solve various COPs, in which a low-latency COP model is enabled by the differentiable ranking (DiffRank)-based distillation and it can be boosted by Gumbel trick-based efficient sequence sampling.
 \item We test the framework with well-known COPs in system areas such as priority-based task scheduling in real-time systems and multidimensional knapsack in resource management, demonstrating the robustness of our framework approach to various problem conditions.
\end{itemize}



\section{Our Approach} \label{sec:framework} 

In this section, we describe the overall structure of the RLRD framework with two building blocks, the deep RL-based COP model structure, and the ranking distillation procedure.

\begin{figure}[t]
    \centering
    \includegraphics[width=0.46\textwidth]{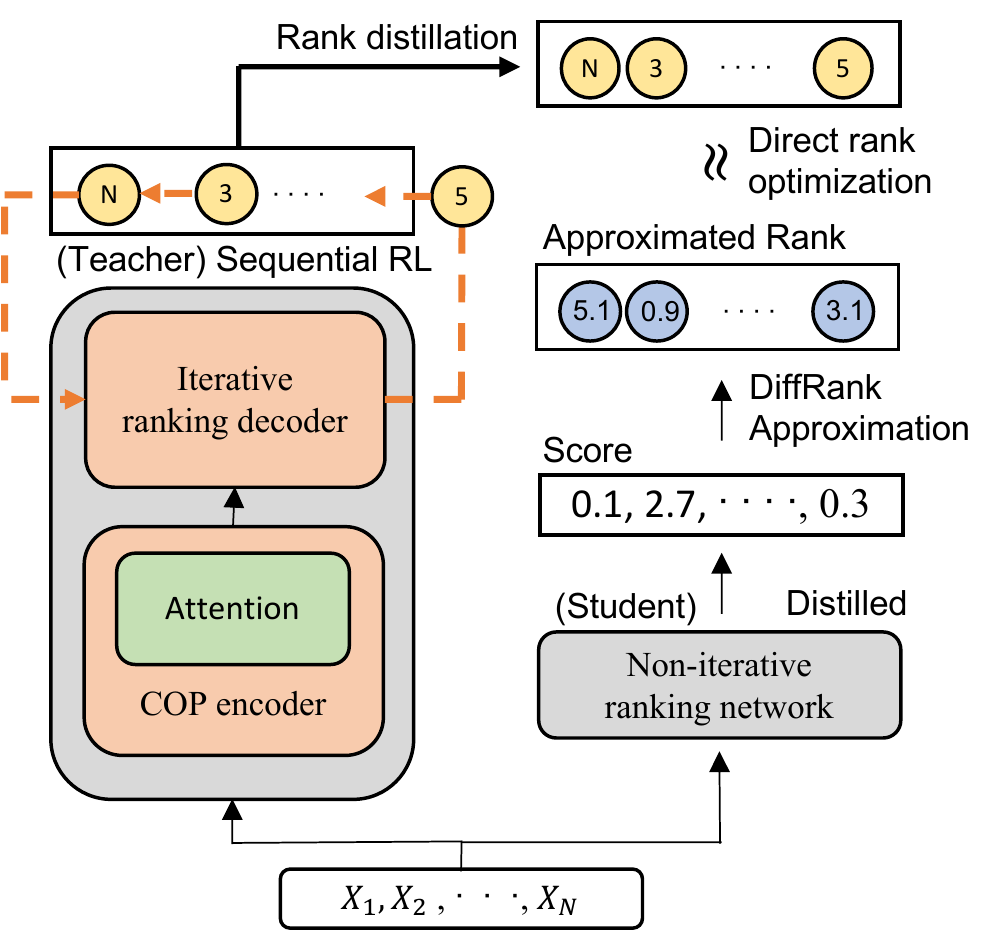}
    \caption{Overview of RLRD framework}
    \label{fig:framework}
\end{figure}

\subsection{Framework Structure} \label{subsec:structure} 

In general, retraining or fine-tuning is needed to adapt a deep learning-based COP model for varying conditions on production system requirements. The RLRD framework supports such model adaptation through knowledge distillation. As shown in Figure~\ref{fig:framework}, (1) for a COP, a learning-to-rank (teacher) policy in the encoder-decoder model is first trained by sequential RL, and (2) it is then transferred through the DiffRank-based distillation to a student model with non-iterative ranking operations, according to a given deployment configuration, e.g., requirements on low-latency inference or model size. For instance, a scheduling policy is established by RL to make inferences on the priority order of a set of tasks running on a real-time multiprocessor platform, and then it is distilled into a low-latency model to make same inferences with some stringent delay requirement.

\subsection{Reinforcement Learning-to-Rank}
Here, we describe the encoder-decoder structure of our RL-to-Rank model for COPs, and explain how to train it. 
Our teacher model is based on a widely adopted attentive  structure~\cite{kool2018attention}. 
In our model representation, we consider parameters $\theta$ (e.g., $W_{\theta}x + b_{\theta}$ for Affine transformation of vector $x$), and we often omit them for simplicity. 
In an RL-to-Rank model, an encoder takes the features of $N$-items as input, producing the embeddings for the $N$-items, and a decoder conducts ranking decisions iteratively on the embeddings, yielding a permutation for the $N$-items. This encoder-decoder model is end-to-end trained by RL.     

\subsubsection{COP Encoder.}\label{subsubsec:encoder}
In the encoder, each item $x_i \in \mathbb{R}^{d}$ containing $d$ features is first converted into vector $h_i^{(0)}$ through the simple Affine transformation, 
    $h_i^{(0)} = Wx_i + b.$
Then, for $N$-items, $(h \times N)$-matrix, $H^{(0)} = [h_1^{(0)}, \cdots, h_N^{(0)}]$ is passed into the $L$-attention layers, where 
each attention layer consists of a Multi-Head Attention layer (MHA) and a Feed Forward network (FF). Each sub-layer is computed with skip connection and Batch Normalization (BN). 
For $l \in \{1, \cdots, L\}$, $H^{(l)}$ are updated by
\begin{equation}
\begin{split}
    H^{(l)} &= \textbf{BN}^{(l)}(X + \textbf{FF}^{(l)}(X)),
    \\ 
    X &= \textbf{BN}^{(l)}(H^{(l-1)} + \textbf{MHA}(H^{(l-1)}))
\end{split}
\label{equ:embedding}
\end{equation}
where
\begin{equation}
    \textbf{MHA}(X) = W_G(\textbf{AM}_1(X) \odot  \cdots \odot \textbf{AM}_{d_h}(X)),
\end{equation}
$\odot$ is the concatenation of tensors, $d_h$ is a fixed positive integer, and $W_G$ is a learnable parameter. \textbf{AM} is given by
\begin{equation}
    \textbf{AM}_j(X) =  W_V(X) \textbf{Softmax}(\frac{1}{d_h}(W_K(X)^TW_Q(X)))
\end{equation}
where $W_Q, W_K$ and $W_V$ denote the layer-wise parameters for query, key and value~\cite{vaswani2017attention}. The result output $H^{(L)}$ in~\eqref{equ:embedding} is the embedding for the input $N$-items, which are used as input to a decoder in the following.

\subsubsection{Ranking Decoder.}\label{sub:decoder}
With the embeddings $H^{(L)}$ for $N$-items from the encoder, the decoder sequentially selects items to obtain an $N$-sized permutation $\phi = [\phi_1, \phi_2, \dots, \phi_N]$ where distinct integers $\phi_i \in \{1,2,\dots,N\}$ correspond to the indices of the $N$-items. That is, item $\phi_1$ is selected first, so it is assigned the highest ranking (priority), and item $\phi_2$ is assigned the second, and so on. 
Specifically, the decoder establishes a function to rank $N$-items stochastically, 
\begin{equation}
    \mathbb{P}(\phi \vert H^{(L)}) = \prod_{t=2}^{N} \mathbb{P} (\phi_t | \phi_1, \dots, \phi_{t-1}, H^{(L)})
\label{equ:probpa}
\end{equation} 
where $\mathbb{P} (\phi_t | \phi_1, \dots, \phi_{t-1}, H^{(L)})$ represents a probability that item $\phi_t$ is assigned the $t$ th rank. 

From an RL formulation perspective, in~\eqref{equ:probpa}, the information about $N$-items ($H^{(L)}$) including ranking-assigned items [$\phi_1, \dots, \phi_{t-1}$] until $t$ corresponds to \textit{state} $o_t$, and selecting $\phi_t$ corresponds to \textit{action} $a_t$. That is, a state contains a partial solution over all permutations and an action is a one-step inference to determine a next ranked item. 
Accordingly, the stochastic ranking function above can be rewritten as $\theta$-parameterized policy $\pi_{\theta}$ for each timestep $t$.
\begin{equation}
    \pi_{\theta}(o_t, a_t) = \mathbb{P}(a_t \mid o_t) = \mathbb{P} (\phi_t | \phi_1, \dots, \phi_{t-1}, H^{(L)})
\label{equ:policy}    
\end{equation}
This policy is learned based on problem-specific reward signals.
%
To establish such policy $\pi_{\theta}$ via RL, we formulate a learnable score function of item $x_i$ upon state $o_t$, which is used to estimate $\mathbb{P}(a_t = i \mid o_t)$, e.g., 
\begin{equation}
\begin{split}
 & \textbf{SCORE}_{H^{(L)}, l^{(t-1)}}(x_i) \\=
	& \begin{cases}
	  10 \, * \textbf{tanh}\left(\textbf{Att}(x^{(g)}, H^{(L)})_i \right) & \text{if} \ i \notin [\phi_1, \cdots, \phi_{t-1})	 \\
	  -\infty &\text{otherwise}
	\end{cases}
\end{split}
\label{eq:score}
\end{equation}
where $l^{(t-1)}$ is the embedding of an item ($x_{\phi_{t-1}}$) selected at timestep $t-1$, and 
 $x^{(g)}$ is a global vector obtained by
\begin{equation}
 x^{(g)} = l^{(t-1)} \odot \textbf{Att}(l^{(t-1)}, H^{(L)}).
\end{equation}
%
%
%
Note that $l^0$ is randomly initialized.
To incorporate the alignment between $l^{(t-1)}$ and $H^{(L)}$ in $x^{(g)}$, we use Attention~\cite{vaswani2017attention},
\begin{equation}
  \begin{split}
  \textbf{Att}(q, Y) = \sum _{i=1}^N \alpha_i y_i, \ \ \alpha_i &= \textbf{Softmax}(w_1, \cdots, w_N), \\
  w_i &= V_A \ \textbf{tanh} \, W_A[y_i \odot q ]   
  \end{split}
\end{equation}
for query $q$ and vectors $Y = [y_1, \cdots, y_N]$, 
where $V_A$ and $W_A$ are learnable parameters. 
Finally, we have the policy $\pi_\theta$ that calculates the ranking probability that the $i$th item is selected next upon state $o_t$. 
\begin{equation}
	\pi_\theta(o_t, a_t = i) = \mathbb{P}(a_t=i \mid o_t) = \frac{e^{\textbf{SCORE}(x_i)}}{\sum_{k=1}^N e^{\textbf{SCORE}(x_k)}}.
\label{condprob}
\end{equation}

\subsubsection{Training.}
For end-to-end training the encoder-decoder, we use the REINFORCE algorithm~\cite{williams1992simple}, which is effective for episodic tasks, e.g., problems formulated as ranking $N$-items. Suppose that for a problem of $N$-items, we obtain 
an episode with  
\begin{equation}
 T^{(\theta)} = (s_1, a_1, r_1, \cdots, s_N, a_N, r_N)
\end{equation}
that are acquired by policy $\pi_{\theta}$, where $s, a$ and $r$ are state, action and reward samples.
We set the goal of model training to maximize the expected total reward by $\pi_{\theta}$,
\begin{equation}
\label{totalreturn}
J(\theta) = \mathbb{E}_{\pi_\theta} \left( \sum _{t=1}^N \gamma^t r_t \right)
\end{equation}
where $\gamma$ is a discount rate, and use the policy gradient ascent.
\begin{equation}
   \theta \leftarrow \theta + \lambda \sum_{t=1}^N \nabla_\theta \log\pi_\theta(o_t, a_t)\left(G_t - b(t)\right) 
\end{equation}
Note that $G_t = \sum_{k=t}^N \gamma^{k-t} r_{k+1}$ is a return, $\lambda$ is a learning rate, and $b(t)$ is a baseline used to accelerate the convergence of model training. 

\subsection{Learning-to-Rank Distillation} \label{subsec:distillation} 

In the RLRD framework, the ranking decoder repeats $N$-times of selection to rank $N$-items through its sequential structure. While the decoder structure is intended to extract the relational features of items that have not been selected, the high computing complexity of iterative decoding renders difficulties in the application of the framework to mission-critical systems. 
To enable fast inferences without significant degradation in model performance, we employ knowledge distillation from an RL-to-rank model with iterative decoding to a simpler model. Specifically, we use a non-iterative, score-based ranking model as a student in knowledge distillation, which takes the features of $N$-items as input and directly produces a score vector for the $N$-items as output. A score vector is used to rank the $N$-items. 

For $N$-items, the RL-to-rank model produces ranking vector as supervised label $\textbf{y} = [\phi_1, \phi_2, \dots, \phi_N]$, and by distillation, the student model learns to produce such a score vector $\textbf{s}$ maximizing the similarity between $\textbf{y}$ and the corresponding ranking of $\textbf{s}$, say $\textbf{rank}(\textbf{s})$. 
For example, given a score vector $\textbf{s} = [2.4, 1.3, 3.0, 0.1]$ for $4$-items, it is sorted to $[3.0, 2.4, 1.3, 0.1]$, so $\textbf{rank}(\textbf{s}) = [2, 3, 1, 4]$. The ranking distillation loss is defined as
\begin{equation}
	\mathcal{L}(\textbf{y}, \textbf{s}) = \mathcal{L}_R\left(\textbf{y}, \textbf{rank}(\textbf{s})\right)    
	\label{distillloss}
\end{equation}
where $\mathcal{L}_R$ is a differentiable evaluation metric for the similarity of two ranking vectors. We use mean squared error (MSE) for $\mathcal{L}_R$, because minimizing MSE of two ranking vectors is equivalent to maximizing the Spearman-rho correlation of two rankings \textbf{y} and \textbf{rank}(\textbf{s}).

\subsubsection{Differentiable Approximated Ranking.}
To distill with the loss in~\eqref{distillloss} using gradient descent, the ranking function $\textbf{rank}$ needs to be differentiable with non-vanishing gradient. However, differentiating $\textbf{rank}$ has a problem of vanishing gradient because a slight shift of score $\textbf{s}$ does not usually affect the corresponding ranking. Thus, we revise the loss in~\eqref{distillloss} using an approximated ranking function having nonzero gradients in the same way of~\cite{blondel2020fast}. 

Consider score $\textbf{s} = [s_1, \cdots, s_N]$ and $N$-permutation $\phi$ which is a bijection from $\{1, \cdots, N\}$ to itself.
A descending sorted list of $\textbf{s}$ is represented as 
\begin{equation}
    \textbf{s}_{\phi} = [s_{\phi(1)}, \cdots, s_{\phi(N)}]
    \label{equ:sort}
\end{equation}
where $s_{\phi(1)} \geq \cdots \geq s_{\phi(N)}$. 
Accordingly, the ranking function \textbf{rank} : $\mathbb{R}^N \rightarrow \mathbb{R}^N$ is formalized as 
\begin{equation}
	\textbf{rank}(\textbf{s}) = [\phi^{-1}(1), \cdots, \phi^{-1}(N)]
\label{equ:rkdef}
\end{equation}
where $\phi^{-1}$ is an inverse of $\phi$, which is also a permutation. For example, consider $\textbf{s} = [2.4, 1.3, 3.0, 0.1]$. Its descending sort is $[3.0, 2.4, 1.3, 0.1]$, so we have $\phi(1) = 3, \phi(2) = 1, \phi(3) = 2$ and $\phi(4) = 4$. Accordingly, we have  $\textbf{rank}(\textbf{s}) = [2, 3, 1, 4]$. 

To implement the DiffRank, a function $\tilde{\textbf{rk}}_{\epsilon}$ is used, which
approximates \textbf{rank} in a differential way with nonzero gradients such as
\begin{equation}
	\tilde{\textbf{rk}}_{\epsilon}(\textbf{s}) = \text{ArgMin}\left\{\frac{1}{2}\lVert\textbf{x} + \frac{\textbf{s}}{\epsilon}^2 \rVert \mid \textbf{x} \in \mathcal{P}(\tau)\right\}
	\label{equ:apprank}.
\end{equation}
Here $\mathcal{P}(\tau)$ is called a perumutahedron, which is a convex hull generated by the permutation $\tau = [N, N-1, \cdots, 1]$ with $\epsilon > 0$. 
As explained in~\cite{blondel2020fast}, the function $\tilde{\textbf{rk}}_{\epsilon}$ converges to $\textbf{rank}$ as $\epsilon \rightarrow 0$, while it always preserves the order of $\textbf{rank}(\textbf{s})$. That is, given $\textbf{\textbf{s}}_{\phi}$ in~\eqref{equ:sort} and  $\tilde{\textbf{rk}}_{\epsilon}(s) = [\psi_1, \cdots, \psi_N]$, we have $\psi_{\phi(1)} \leq \cdots \leq \psi_{\phi(N)}$.

In addition, we also consider a problem-specific loss. For example, in the knapsack problem, an entire set of items can be partitioned into two groups, one for selected items and the other for not selected items. 
We can penalize the difference of partitions obtained from label \textbf{y} and target output score $\textbf{s}$ by the function $\mathcal{L}_P$. Finally the total loss is given by 
\begin{equation}
    \mathcal{L} (\textbf{y}, \textbf{s}) = \alpha \, \mathcal{L}_R \left( \textbf{y}, \tilde{\textbf{rk}}_{\epsilon}(\textbf{s}) \right) + (1-\alpha) \mathcal{L}_P(\textbf{y}, \textbf{s})
    \label{equ:finalloss}
\end{equation}
where $\alpha \in [0, 1]$. 
%
The overall distillation procedure is illustrated in Algorithm~\ref{alg:distill}.

Here, we present the explicit nonvanishing gradient form of our ranking loss function $\mathcal{L}_R$, 
where its proof can be found in 
Appendix A. 
%
\begin{prop}
	Fix $\textbf{r} = [r_1, \cdots, r_n] \in \mathbb{R}^n$. Let $\tilde{\textbf{rk}}_{\epsilon} : \mathbb{R}^n \longrightarrow \mathbb{R}^n$ as in~\eqref{equ:apprank} and $L : \mathbb{R}^n \longrightarrow \mathbb{R}$ where $L(\textbf{y}) = \frac{1}{2} \lVert \textbf{y} - \textbf{r} \rVert _2^2$. Let
	$g = L \circ \tilde{\textbf{rk}}_\epsilon$,
	\begin{align}
		\tilde{\textbf{rk}}_{\epsilon}(\textbf{s}) = [\tilde{r}_1, \cdots, \tilde{r}_n],
	\end{align} and $e_i = \tilde{r}_i - r_i$.
	Then, we have
	\begin{align}
	\label{propmtx}
		\frac{\partial g}{\partial \textbf{x}}(\textbf{s}) & = -\frac{\textbf{I}}{\epsilon} [e_1 , \cdots, e_n] \\ & @ \left(\textbf{I} - 
		\begin{bmatrix}
			\frac{1}{k_1}\textbf{1}_{k_1} & \textbf{0} & \dots & \textbf{0}	\\
			\textbf{0} & \frac{1}{k_2}\textbf{1}_{k_2} & \dots & \textbf{0}	\\
			\vdots 		&	\vdots	&	\vdots	&	\vdots	\\
			\textbf{0}	&	\textbf{0}	&	\dots	&	\frac{1}{k_m}\textbf{1}_{k_m}
		\end{bmatrix} \right)_{\phi}
	\end{align}
	where @ is a matrix multiplication, $k_1 + \cdots + k_m = n$, $\textbf{1}_{k_j}$ is a square matrix whose entries are all $1$ with size $k_j \times k_j$, and $\phi$ is an $n$-permutation. Here, for any matrix $M$, $M_{\phi}$ denotes row and column permutation of $M$ according to $\phi$.
\end{prop}

\begin{algorithm}[t]
	\caption{Learning-to-rank distillation}
	\begin{algorithmic}[1]
	\fontsize{8pt}{12pt}\selectfont
	\STATE Load (teacher) RL policy $\pi$
	\STATE Initialize parameter $\theta_s$ of student model
	\STATE \textbf{Input} : Sample batch $\mathcal{B}$ and learning rate $\lambda_s$.
	\FOR{$1\sim N_s$}
		\STATE $\nabla \theta_s \longleftarrow 0$.
		\FOR{Itemset $I \in \mathcal{B}$}
			\STATE Get rank  = $\{\phi(1), \cdots, \phi(n)\}$ from $\pi$
			\STATE Get score $\textbf{s} = \{s_1 ,\cdots, s_n\}$ from target model.
			\STATE Get approx. ranking $\tilde{\textbf{rk}_\epsilon}(\textbf{s}) = \{\tilde{\psi}(1), \cdots, \tilde{\psi}(n)\}$ using (\ref{equ:apprank})
			\STATE Calculate loss $J(\theta_s) = \mathcal{L}(L, \tilde{R})$ using~\eqref{equ:finalloss}
			\STATE $\nabla \theta_s \longleftarrow \nabla \theta_s + \nabla J(\theta_s)$
		\ENDFOR
		\STATE $\theta_s \longleftarrow \theta_s + \lambda_s \nabla \theta_s$
	\ENDFOR
	\end{algorithmic}
	\label{alg:distill}
\end{algorithm}

\subsubsection{Efficient Sequence Sampling.}
As explained, we use a score vector in~\eqref{equ:sort} to obtain its corresponding ranking vector deterministically. On the other hand, if we treat such a score vector as an un-normalized log-probability of a categorical distribution on $N$-items, we can randomly sample from the distribution using the score vector \textit{without replacement} and obtain a ranking vector for the $N$-items. Here, the condition of \textit{without replacement} specifies that the distribution is renormalized so that it sums up to $1.0$ for each time to sample an item. This $N$-times drawing and normalization increases the inference time. Therefore, to compute rankings rapidly, we exploit Gumbel trick~\cite{gumbel1954maxima,maddison2014sampling}.


Given score $\textbf{s}$, consider the random variable $\textbf{S} = \textbf{s}+Z$ where 
\begin{equation}
  Z = -\log\left(-\log(\textbf{Unif}(0,1))\right), 
\end{equation}
and suppose that $\textbf{S}$ is sorted by $\phi$, as in \eqref{equ:sort}. Note that   $\phi(1), \phi(2), \cdots, \phi(N)$ are random variables.
\begin{theorem}~Appendix A in \cite{kool2019estimating}. \label{theorem:gumbelperm}
    For each k $\in \{1, 2, \cdots, N\}$ ,
    \begin{equation}
    	\mathbb{P}(\phi(1) = i_1, \cdots, \phi(k) = i_k) = \prod _{j=1}^k \frac{\textnormal{exp} (s_{i_l})}{\sum _{l \in R_j^{*}} \textnormal{exp} (s_{l})}
    \end{equation}
    where $R_j^{*} = \{1, 2, \cdots, N\} - \{i_1, \cdots, i_{j-1}\}$.
\end{theorem}
This sampling method reduces the complexity to obtain each ranking vector instance from quadratic (sequentially sampling each of $N$-items on a categorical distribution) to log-linear (sorting a perturbed list) of the number of items $N$, improving the efficiency of our model significantly.

\begin{table*}[t]
	\centering
	\begin{adjustbox}{width=1.0\textwidth}
	\begin{tabular}{|c|c|c|c|cc|cc|cc|cc|cc|cc|}
		\hline
		\multirow{2}{*}{$N$} & \multirow{2}{*}{$k$} & \multirow{2}{*}{$w$} & \multirow{2}{*}{$\alpha$} 
		& \multicolumn{2}{c|}{GLOP} & \multicolumn{2}{c|}{Greedy} & \multicolumn{2}{c|}{RL} & \multicolumn{2}{c|}{RL-S}  & \multicolumn{2}{c|}{RD}  & \multicolumn{2}{c|}{RD-G}  \cr
		&&&& - & Time &  Gap & Time &  Gap & Time &  Gap & Time &  Gap & Time & Gap & Time \cr  \hline 
		\multirow{4}{*}{50} & 
		\multirow{4}{*}{3}	&	
		\multirow{2}{*}{200}	&0
					&100\%&0.0060s&97.9\%&0.0003s&99.7\%&0.0706s&98.8\%&2.0051s&97.5\%&0.0029s&100.1\%&0.0152s \cr
		&&&0.9      &100\%&0.0063s&87.6\%&0.0003s&100.2\%&0.0675s&104.5\%&1.8844s&97.7\%&0.0030s&102.2\%&0.0154s \cr
		&&\multirow{2}{*}{500}
		&0			&100\%&0.0066s&97.8\%&0.0003s&99.4\%&0.0686s&99.6\%&1.9768s&97.4\%&0.0029s&99.7\%&0.0159s \cr
		&&&0.9		&100\%&0.0064s&81.4\%&0.0004s&101.5\%&0.0687s&105.4\%&1.8840s&97.9\%&0.0030s&101.5\%&0.0150s \cr
		
		\hline
		\multirow{4}{*}{100} &
		\multirow{4}{*}{10} &
		\multirow{2}{*}{200} & 0 &100\%&0.0950s&101.3\%&0.0005s&102.2\%&0.4444s&101.9\%&12.4996s&100.5\%&0.0060s&102.7\%&0.0198s \cr
		&&&0.9					&100\%&0.0435s&93.4\%&0.0004s&103.2\%&0.4443s&106.9\%&12.3932s&99.2\%&0.0086s&102.6\%&0.0222s	 \cr
		&&
		\multirow{2}{*}{500}&0	&100\%&0.1046s&100.9\%&0.0005s&100.6\%&0.4363s&101.1\%&12.3392s&98.9\%&0.0088s&101.6\%&0.0214s		\cr
		&&&0.9					&100\%&0.0436s&90.2\%&0.0004s&103.5\%&0.4381s&107.1\%&12.3211s&100.4\%&0.0059s&104.6\%&0.0198s		\cr
		
		\hline	
		
		\multirow{4}{*}{150} &
		\multirow{4}{*}{15}	&
		\multirow{2}{*}{200} & 0 &100\%&0.2494s&102.0\%&0.0007s&102.9\%&0.6370s&102.8\%&17.8975s&100.7\%&0.0090s&102.5\%&0.0328s \cr
		&&&0.9					 &100\%&0.0885s&96.7\%&0.0006s&103.4\%&0.5380s&106.4\%&16.2406s&99.7\%&0.0090s&102.3\%&0.0290s \cr
		&&\multirow{2}{*}{500}&0 &100\%&0.2497s&101.8\%&0.0007s&99.1\%&0.6488s&100.1\%&19.4974s&96.6\%&0.0093s&98.6\%&0.0244s \cr
		&&&0.9					 &100\%&0.0425s&92.5\%&0.0006s&103.7\%&0.5289s&107.0\%&14.0924s&100.7\%&0.0088s&103.9\%&0.0292s \cr
		\hline
		
	\end{tabular}
	\end{adjustbox}
	\caption{The Evaluation of MDKP. For each method, Gap denotes the performance (average achieved value) ratio of the method to GLOP, and Time denotes the average inference time for a problem instance. $N$ and $k$ denote the number of items and the size of knapsack resource dimensions, respectively. $w$ denotes the sampling range of item weight on [$1$, $w$], and $\alpha$ denotes the correlation of weight and value of items. The performance is averaged for a testing dataset of 500 item sets.}
	\label{mdkptable}
\end{table*}

\section{Experiments} \label{sec:eval} 

In this section, we evaluate our framework with 
multidimensional knapsack problem (MDKP) and
global fixed-priority task scheduling (GFPS)~\cite{davis2016review}. The problem details including RL formulation, data generation, and model training can be found in 
Appendix B and C. 

\subsection{Multidimensional Knapsack Problem (MDKP)} \label{subsec:eval_knapsack} 

Given values and $k$-dimensional weights of $N$-items, in MDKP, each item is either selected or not for a knapsack with $k$-dimensional capacities to get the maximum total value of selected items. 

For evaluation, we use the performance (the achieved value in a knapsack) by GLOP implemented in the OR-tools~\cite{ortools} as a baseline. 
We compare several models in our framework. 
RL is the RL-to-rank teacher model, and RD is the distilled student model. 
RL-S is the RL model variant with sampling, and RD-G is the RD model variant with Gumbel trick-based sequence sampling. 
In RL-S, the one-step action in~\eqref{condprob} is conducted stochastically, while in RL, it is conducted greedily. 
For both RL-S and RD-G, we set the number of ranking sampling to 30 for each item set, and 
report the the highest score among samples.
In addition, we test the \textit{Greedy} method that exploits the mean ratio of item weight and value. 
 

\subsubsection{Model Performance.}

Table~\ref{mdkptable} shows the performance in MDKP, where GAP denotes the performance ratio to the baseline GLOP and Time denotes the inference time. 
\begin{itemize}
    \item RD shows comparable performance to its respective high-performance teacher RL, with insignificant degradation of average 2.6\% for all the cases. More importantly, RD achieves efficient, low-latency inferences, e.g., 23 and 65 times faster inferences than RL for $N$=50 and $N$=150 cases, respectively.  
    \item RD-G outperforms RL by 0.3\% on average and also achieves 4.4 and 20 times faster inferences than RL for $N$=50 and $N$=150 cases, respectively. Moreover, RD-G shows 2\% higher performance than RD, while its inference time is increased by 3.7 times. 
    \item RL-S shows 1.8\% higher performance than RL model. However, unlike RD-G, the inference time of RL-S is increased linearly to the number of ranking samples (i.e., about 30 times increase for 30 samples).  
    \item As $N$ increases, all methods shows longer inference time, but the increment gap of GLOP is much larger than RL and RD. For example, as $N$ increases from 50 to 150 when $\alpha$=0, the inference time of GLOP is increased by 39 times, while RL and RD shows 9.3 and 3.1 times increments, respectively.
    \item The performance of Greedy degrades drastically in the case of $\alpha$=0.9. This is because the weight-value ratio for items becomes less useful when the correlation is high. Unlike Greedy, our models show stable performance for both high and low correlation cases. 
\end{itemize}

\subsection{Priority Assignment Problem for GFPS} \label{subsec:eval_task} 

\begin{table*}[t]
	\centering
	\begin{adjustbox}{width=0.94\textwidth}
	\begin{tabular}{|c|c|c|cc|cc|cc|cc|cc|}
		\hline
		\multirow{2}{*}{m} & \multirow{2}{*}{N} & \multirow{2}{*}{Util} & \multicolumn{2}{c|}{OPA} & 
		\multicolumn{2}{c|}{RL} & \multicolumn{2}{c|}{RL-S} & \multicolumn{2}{c|}{RD} & \multicolumn{2}{c|}{RD-G}  \cr
		&&& Ratio & Time & Ratio & Time & Ratio & Time & Ratio & Time & Ratio & Time \cr \hline
		
		\multirow{4}{*}{4} & \multirow{4}{*}{32} & 3.0 & 78.1\% & 0.3531s & 87.5\% & 0.0616s & 89.4\% & 0.0697s & 86.5\%& 0.0145s & 90.1\% & 0.0263s \cr
		&& 3.1 &63.5\%&0.3592s&74.8\%&0.0599s&77.6\%&0.0840s&73.8\%&0.0139s & 78.9\% & 0.0390s \cr
	    && 3.2 &44.9\%&0.3487s&56.9\%&0.0623s&60.1\%&0.0991s&56.0\%&0.0140s & 61.2\% & 0.0509s \cr
		&&3.3	& 26.6\% &0.3528s&35.7\%&0.0621s&38.5\%&0.9123s&35.8\%&0.0131s&39.2\%&0.0620s \cr
		\hline
		\multirow{4}{*}{6} & \multirow{4}{*}{48}&
		4.4		&84.2\%&0.4701s&92.5\%&0.1021s&94.3\%&0.1153s&91.76\%&0.0298s&94.3\%&0.0406s \cr
		&& 4.6		&61.9\%&0.4600s&78.4\%&0.1057s&78.4\%&0.1508s&74.6\%&0.0308s&79.3\%&0.0728s \cr
		&& 4.8		&33.2\%&0.4287s&46.5\%&0.1082s&50.4\%&0.1967s&45.4\%&0.0290s&50.8\%&0.1123s  \cr
		&& 5.0		&11.5\%&0.3888s&15.7\%&0.1010s&18.1\%&0.2474s&18.0\%&0.0256s&20.2\%&0.1615s	 \cr
		\hline
		
		\multirow{4}{*}{8} & \multirow{4}{*}{64} &
		5.7		& 92.9\% & 0.6686s & 97.8\% & 0.1437s & 98.6\% & 0.1596s & 97.5\% & 0.0502s & 98.5\% & 0.0537s  \cr
		&& 6.0		&72.9\% & 0.6460s & 86.5\% & 0.1490s & 89.9\% & 0.2043s & 85.0\% & 0.0364s & 88.7\% & 0.0907s \cr
		&& 6.3		&37.6\% & 0.5798s & 53.5\% & 0.1559s & 57.5\% & 0.2800s & 52.5\% & 0.0509s & 57.7\% & 0.1695s  \cr
		&& 6.6		&10.4\%&0.4806s&15.1\%&0.1488s&17.7\%&0.4093s&17.0\%&0.0390s&19.6\%&0.2715s \cr
		\hline

	\end{tabular}
		
	\end{adjustbox}
			\caption{The Evaluation of GFPS. For each method, Ratio denotes the performance in the schedulability ratio $( \frac{\text{num of schedulable task sets}}{\text{num. of task sets}})$, and Time denotes the average inference time for a problem instance. 
			$m$ and $N$ denote the number of processors and the number of tasks, respectively, i.e., scheduling $N$-tasks on an $m$-processor platform. Util denotes the task set utilization, i.e., the sum of task utilization $(\sum \frac{\text{task exe. time}}{\text{task period}})$. The performance is averaged for a testing dataset of 5,000 task sets.}
			\label{gfpstable}
\end{table*}

For a set of $N$-periodic tasks, in GFPS, each task is assigned a priority (an integer from $1$ to $N$) to be scheduled. GFPS with a priority order (or a ranking list) can schedule the $m$ highest-priority tasks in each time slot upon a platform comprised of $m$-processors, with the goal of not incurring any deadline violation of the periodic tasks. 

For evaluation, we need to choose a schedulability test for GFPS, that determines whether a task set is deemed schedulable by GFPS with a priority order. We target a schedulability test called RTA-LC~\cite{Guan2009, Davis2011} which has been known to perform superior to the others in terms of covering schedulable task sets.
We compare our models with Audsley's Optimal Priority Assignment (OPA)~\cite{Audsley1991, Audsley2001} with the state-of-the-art OPA-compatible DA-LC test~\cite{Davis2011}, which is known to have the highest performance compared to other heuristic algorithms. Same as those in MDKP, we denote our models as RL, RL-S, RD, and RD-G. For both RL-S and RD-G, we limit the number of ranking samples to 10 for each task set. 

\subsubsection{Model Performance.}


  

   

   


Table ~\ref{gfpstable} shows the performance in the schedulability ratio of GFPS with respect to different task set utilization settings on an $m$-processor platform and $N$-tasks. 
\begin{itemize}
    \item Our models all show better performance than OPA, indicating the superiority of the RLRD framework. The performance gap is relatively large on the intermediate utilization ranges, because those ranges can provide more opportunities to optimize with a better strategy. For example, when $m$=8, $N$=64 and Util=6.3, RL and RD show 15.9\% and 11.9\% higher schedulability ratio than OPA, respectively, while when $m$=8, $N$=64 and Util=5.7, their gain is 4.9\% and 4.6\%, respectively. 
    \item The performance difference of RD and its teacher RL is about 1\% on average, while the inference time of RD is decreased (improved) by 3.8 times. This clarifies the benefit of the ranking distillation. 
    \item As the utilization (Util) increases, the inference time of RL-S and RD-G becomes longer, due to multiple executions of the schedulability test up to the predefined limit (i.e., 10 times). On the other hand, the inference time of OPA decreases for large utilization; the loop of OPA is terminated when a task cannot satisfy its deadline with the assumption that other priority-unassigned tasks have higher priorities than that task.
    \item  RD-G shows comparable performance to, and often achieves slight higher performance than RL-S. This is the opposite pattern of MDKP where RL-S achieves the best performance. While direct comparison is not much valid due to different sampling methods, we notice the possibility that a distilled student can perform better than its teacher for some cases, and the similar patterns are observed in~\cite{tangrankdistil, kim2016sequence}. 
    %
\end{itemize}

\subsection{Analysis on Distillation} \label{subsec:eval_analysis}

\subsubsection{Effects of Iterative Decoding.}
%
To verify the feasibility of distillation from sequential RL to a score-based ranking model, we measure the difference of the outputs by iterative decoding and greedy sampling.  
%
In the case when the decoder generates the ranking distribution at timestep $t$ and takes  action $a_t=i$ as in \eqref{condprob}, by masking the $i$th component of the distribution and renormalizing it, we can obtain a renormalized distribution $P_{\text{RE}}$. In addition, consider another probability distribution $P_\text{DE}$ generated by the decoder at $t+1$. 

\begin{figure}[t]
    \centering
    \includegraphics[width=0.46\textwidth]{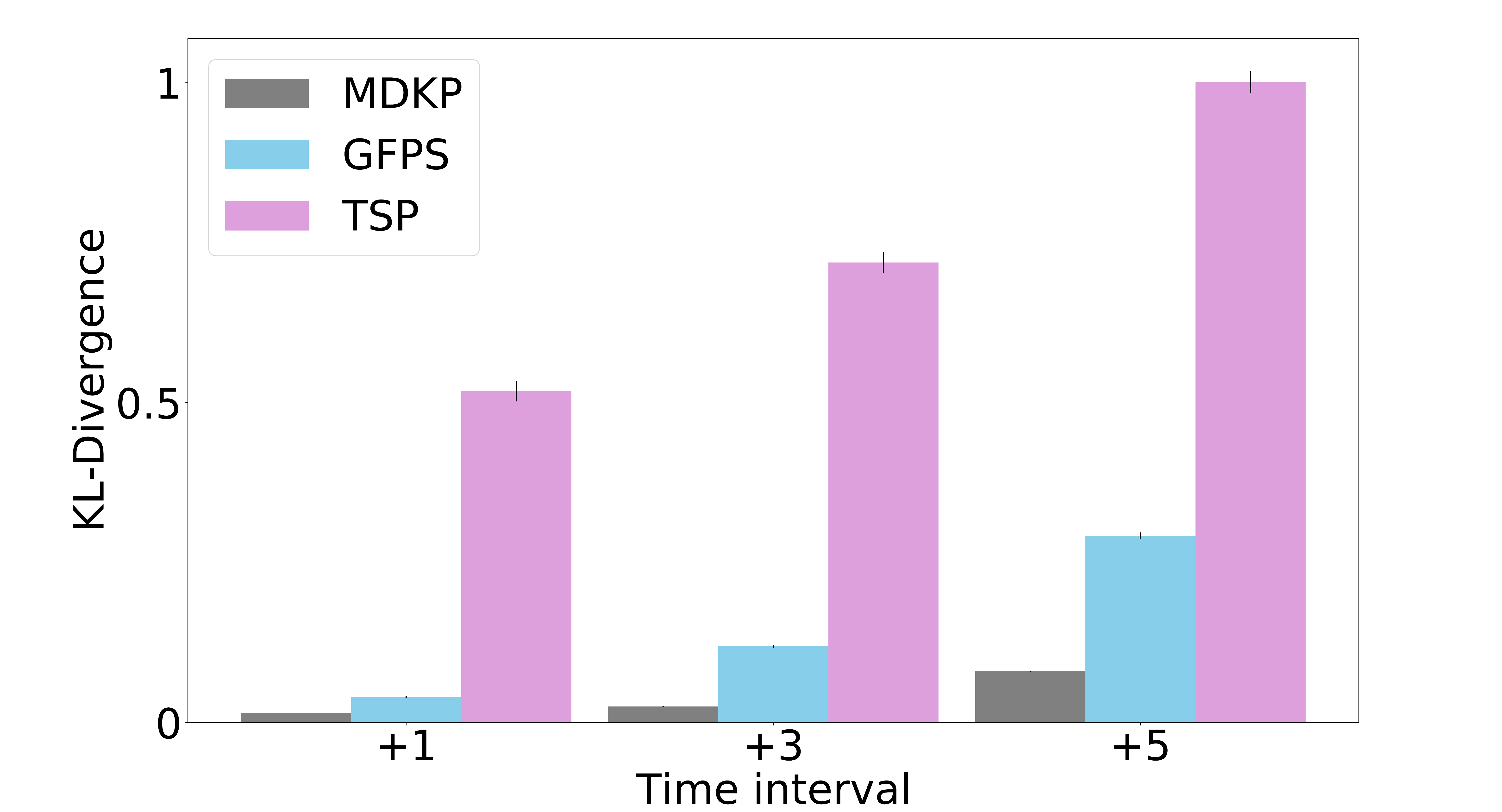}
    \caption{The KL divergence at different timesteps $t$ and $t+\delta$ is used to estimate the difference of decoder driven ranking distribution ($P_{DE}$) and renormalized ranking distribution ($P_{RE}$), where time intervals $\delta \in \{ 1, 3, 5\}$. Scales are normalized.}
    \label{fig:kldiv}
\end{figure}

Figure~\ref{fig:kldiv} illustrates the difference of the distributions in terms of KL-divergence on three specific COPs, previously explained MDKP and GFPS as well as Traveling Salesman Problem (TSP). As shown, MDKP and GFPS maintain a low divergence value, implying that the ranking result of decoding with many iterations can be approximated by decoding with no or few iterations.   
Unlike MDKP and GFPS, TSP shows a large divergence value. This implies that many decoding iterations are required to obtain an optimal path. 
Indeed, in TSP, we obtain good performance by RL (e.g, 2\% better than a heuristic method), but we hardly achieve comparable performance to RL when we test RD. The experiment and performance for TSP can be found in Appendix D.

\subsubsection{Effects of Distillation Loss.} 
To evaluate the effectiveness of DiffRank-based distillation, we implement other ranking metrics such as a \emph{pairwise} metric in~\cite{ranknet} and a \emph{listwise} metric in~\cite{listnet} and test them in the framework as a distillation loss. 

Table~\ref{mdkploss} and~\ref{gfpsloss} show the performance in MDKP and GFPS, respectively, achieved by different distillation loss functions, where RD denotes our distilled model trained with DiffRank-based distillation loss, and the performance of the other two is represented as the ratio to RD. Note that they all use the same RL model as a teacher in this experiment. 

As shown, RD achieves consistently better performance than the others for most cases. Unlike RD, the other methods commonly show data-dependent performance patterns. The pairwise method (with pairwise distillation loss) achieves performance similar to or slightly lower than RD in MDKP but
shows much lower performance than RD in GFPS. The listwise method shows the worst performance for many cases in both MDKP and GFPS except for the cases of $\alpha=0.9$ in MDKP. These results are consistent with the implication in Figure~\ref{fig:kldiv} such that GFPS has larger divergence than MDKP and thus GFPS is more difficult to distill, giving a large performance gain to RD.

\begin{table}[h]
	\centering
	\begin{adjustbox}{width=0.4 \textwidth}
	\begin{tabular}{|c|c|c|c|c|c|c|}
		\hline
		N&k&w&$\alpha$&RD&Pairwise&Listwise \cr \hline
		\multirow{2}{*}{50} & \multirow{2}{*}{5} & \multirow{2}{*}{200} & 0 & 100\% & 99.4\% & 69.8\% \cr
		&&&0.9 &100\%&99.1\%& 98.5\% \cr \hline
		\multirow{2}{*}{100}& \multirow{2}{*}{10} & \multirow{2}{*}{200} & 0 & 100\% & 98.8\% & 80.1\% \cr
		&&&0.9 &100\%&100.1\%& 99.2\% \cr \hline
		\multirow{2}{*}{150}& \multirow{2}{*}{15} & \multirow{2}{*}{200} & 0 & 100\% & 95.4\% & 86.8\%  \cr
		&&&0.9 &100\%&99.6\%&99.6\% \cr \hline
	\end{tabular}
	\end{adjustbox}
	\caption{Ranking Loss Comparison in MDKP. 
	}
	\label{mdkploss}
\end{table}

\begin{table}[h]
	\centering
	\begin{adjustbox}{width=0.4 \textwidth}
	\begin{tabular}{|c|c|c|c|c|c|}
	\hline
	m & N & Util & RD & Pairwise & Listwise \cr	\hline
	\multirow{1}{*}{4} & \multirow{1}{*}{32} & 3.1 &100\%&90.1\%&72.8\% \cr
    \hline
	\multirow{1}{*}{6} & \multirow{1}{*}{48} & 4.6 &100\%&91.3\%&59.6\% \cr
	\hline
	\multirow{1}{*}{8} & \multirow{1}{*}{64} & 6.0 &100\%&91.6\%&49.2\% \cr
	\hline
	\end{tabular}	
	\end{adjustbox}
	\caption{Ranking Loss Comparison in GFPS. 
	}
	\label{gfpsloss}
\end{table}

\color{black}
\section{Related Work} \label{sec:related}

Advanced deep neural networks combined with RL algorithms showed the capability to address various COPs in a data-driven manner with less problem-specific customization. In~\cite{bello2016neural}, the pointer network was introduced to solve TSP and other geometric COPs, and in~\cite{kool2018attention}, a transformer model was incorporated for more generalization. Besides the pointer network, a temporal difference based model showed positive results in the Job-Shop problem~\cite{zhangjobshop}, and deep RL-based approaches such as Q-learning solvers~\cite{kpqlearning} were explored for the knapsack problem~\cite{kpqlearning}. 
Several attempts have been also made to address practical cases formulated in the knapsack problem, e.g., maximizing user engagements under business constraints~\cite{homepagerelevance, emailvolumeoptimize}. 

Particularly, in the field of computer systems and resource management, there have been several works using deep RL to
tackle system optimization under multiple, heterogeneous resource constraints in the form of COPs, e.g., cluster resource management~\cite{mao2016resource,mao2018learning}, compiler optimization~\cite{chen2018tvm}.
While we leverage deep RL techniques to address COPs in the same vein as those prior works, we focus on efficient, low-latency COP models. 

The ranking problems such as prioritizing input items based on some scores have been studied in the field of information retrieval and recommendation systems. 
A neural network based rank optimizer using a pairwise loss function was first introduced in~\cite{ranknet}, and other ranking objective functions were developed to optimize relevant metrics with sophisticated network structures. For example, Bayesian Personalized Ranking~\cite{rendle2012} is known to maximize the AUC score of given item rankings with labeled data.
However, although these approaches can bypass the non-differentiability of ranking operations, the optimization is limited to some predefined objectives such as NDCG or AUC; thus, it is difficult to apply them to COPs because the objectives do not completely align with the COP objectives.
To optimize arbitrary objectives involving nondifferentiable operations such as ranking or sorting, several works focused on smoothing nondifferentiable ranking operations~\cite{grover2019stochastic,blondel2020fast}. They are commonly intended to make arbitrary objectives differentiable by employing relaxed sorting operations. 


Knowledge distillation based on the teacher-student paradigm has been an important topic in machine learning to build a compressed model~\cite{hinton2015distilling} and showed many successful practices in image classification~\cite{pmlr-v139-touvron21a} and natural language processing~\cite{kim2016sequence,jiao-etal-2020-tinybert}. However, knowledge distillation to ranking models has not been fully studied. 
A ranking distillation for recommendation system was introduced in~\cite{tangrankdistil}, and recently, a general distillation framework RankDistil~\cite{Reddi2021RankDistilKD} was presented with several loss functions and optimization schemes specific to top-$K$ ranking problems. These works exploited pairwise  objectives and sampling based heuristics to distill a ranking model, but rarely focused on arbitrary objectives and sequential models, which are required to address various COPs. 
The distillation of sequential models was investigated in several works~\cite{kim2016sequence,jiao-etal-2020-tinybert}. However, to the best of our knowledge, our work is the first to explore the distillation from sequential RL models to score-based ranking models.

\section{Conclusion} \label{sec:conclude} 
In this paper, we presented a distillation-based COP framework by which an efficient model with high-performance is achieved. Through experiments, we demonstrate that it is feasible to distill the ranking policy of deep RL to a score-based ranking model without compromising performance, thereby enabling the low-latency inference on COPs. %

The direction of our future work is to adapt our RL-based encoder-decoder model and distillation procedure for various COPs with different consistency degrees between embeddings and decoding results and to explore meta learning for fast adaptation across different problem conditions.

\section*{Acknowledgement}
We would like to thank anonymous reviewers for their valuable
comments and suggestions.

This work was supported by the Institute for Information and Communications Technology Planning and Evaluation (IITP) under Grant 2021-0-00875 and 2021-0-00900, 
by the ICT Creative Consilience Program supervised by the IITP under Grant IITP-2020-0-01821, 
by Kakao I Research Supporting Program, and by Samsung Electronics.

\bibliography{aaai22.bib}

\begin{thebibliography}{46}
\providecommand{\natexlab}[1]{#1}

\bibitem[{Afshar et~al.(2020)Afshar, Zhang, Firat, and Kaymak}]{kpqlearning}
Afshar, R.~R.; Zhang, Y.; Firat, M.; and Kaymak, U. 2020.
\newblock A State Aggregation Approach for Solving Knapsack Problem with Deep
  Reinforcement Learning.
\newblock In \emph{Proceedings of the 12th Asian Conference on Machine
  Learning}, volume 129, 81--96.

\bibitem[{Agarwal et~al.(2015)Agarwal, Chatterjee, Yang, and
  Zhang}]{homepagerelevance}
Agarwal, D.; Chatterjee, S.; Yang, Y.; and Zhang, L. 2015.
\newblock Constrained Optimization for Homepage Relevance.
\newblock In \emph{Proceedings of the 24th International Conference on World
  Wide Web Companion}, 375--384.

\bibitem[{Ahuja and Orlin(2001)}]{pavalgorithm}
Ahuja, R.~K.; and Orlin, J.~B. 2001.
\newblock A Fast Scaling Algorithm for Minimizing Separable Convex Functions
  Subject to Chain Constraints.
\newblock \emph{Oper. Res.}, 784--789.

\bibitem[{Audsley(1991)}]{Audsley1991}
Audsley, N.~C. 1991.
\newblock \emph{Optimal priority assignment and feasibility of static priority
  tasks with arbitrary start times}.
\newblock Citeseer.

\bibitem[{Audsley(2001)}]{Audsley2001}
Audsley, N.~C. 2001.
\newblock On priority assignment in fixed priority scheduling.
\newblock \emph{Inf. Process. Lett.}, 79(1): 39--44.

\bibitem[{Bello et~al.(2017)Bello, Pham, Le, Norouzi, and
  Bengio}]{bello2016neural}
Bello, I.; Pham, H.; Le, Q.~V.; Norouzi, M.; and Bengio, S. 2017.
\newblock Neural Combinatorial Optimization with Reinforcement Learning.
\newblock In \emph{Proceedings of the 5th International Conference on Learning
  Representations (ICLR)}.

\bibitem[{Blondel et~al.(2020)Blondel, Teboul, Berthet, and
  Djolonga}]{blondel2020fast}
Blondel, M.; Teboul, O.; Berthet, Q.; and Djolonga, J. 2020.
\newblock Fast Differentiable Sorting and Ranking.
\newblock In \emph{Proceedings of the 37th International Conference on Machine
  Learning (ICML)}, volume 119, 950--959.

\bibitem[{Brandenburg and Gul(2016)}]{brandenburg2016global}
Brandenburg, B.~B.; and Gul, M. 2016.
\newblock Global Scheduling Not Required: Simple, Near-Optimal Multiprocessor
  Real-Time Scheduling with Semi-Partitioned Reservations.
\newblock In \emph{Proceedings of the 37th {IEEE} Real-Time Systems Symposium
  (RTSS)}, 99--110.

\bibitem[{Burges et~al.(2005)Burges, Shaked, Renshaw, Lazier, Deeds, Hamilton,
  and Hullender}]{ranknet}
Burges, C. J.~C.; Shaked, T.; Renshaw, E.; Lazier, A.; Deeds, M.; Hamilton, N.;
  and Hullender, G.~N. 2005.
\newblock Learning to rank using gradient descent.
\newblock In \emph{Proceedings of the 22nd International Conference on Machine
  Learning (ICML)}, volume 119, 89--96.

\bibitem[{Cao et~al.(2007)Cao, Qin, Liu, Tsai, and Li}]{listnet}
Cao, Z.; Qin, T.; Liu, T.; Tsai, M.; and Li, H. 2007.
\newblock Learning to rank: from pairwise approach to listwise approach.
\newblock In \emph{Proceedings of the 24th International Conference on Machine
  Learning {(ICML})}, volume 227, 129--136.

\bibitem[{Chen et~al.(2018)Chen, Moreau, Jiang, Zheng, Yan, Shen, Cowan, Wang,
  Hu, Ceze, Guestrin, and Krishnamurthy}]{chen2018tvm}
Chen, T.; Moreau, T.; Jiang, Z.; Zheng, L.; Yan, E.~Q.; Shen, H.; Cowan, M.;
  Wang, L.; Hu, Y.; Ceze, L.; Guestrin, C.; and Krishnamurthy, A. 2018.
\newblock {TVM:} An Automated End-to-End Optimizing Compiler for Deep Learning.
\newblock In \emph{Proceedings of the 13th {USENIX} Symposium on Operating
  Systems Design and Implementation (OSDI)}, 578--594.

\bibitem[{Chu and Beasley(1998)}]{chu1998genetic}
Chu, P.~C.; and Beasley, J.~E. 1998.
\newblock A genetic algorithm for the multidimensional knapsack problem.
\newblock \emph{Journal of heuristics}, 4(1): 63--86.

\bibitem[{Davis and Burns(2011)}]{Davis2011}
Davis, R.~I.; and Burns, A. 2011.
\newblock Improved priority assignment for global fixed priority pre-emptive
  scheduling in multiprocessor real-time systems.
\newblock \emph{Real Time Syst.}, 47(1): 1--40.

\bibitem[{Davis et~al.(2016)Davis, Cucu{-}Grosjean, Bertogna, and
  Burns}]{davis2016review}
Davis, R.~I.; Cucu{-}Grosjean, L.; Bertogna, M.; and Burns, A. 2016.
\newblock A review of priority assignment in real-time systems.
\newblock \emph{J. Syst. Archit.}, 65: 64--82.

\bibitem[{Emberson, Stafford, and Davis(2010)}]{Emberson2010}
Emberson, P.; Stafford, R.; and Davis, R. 2010.
\newblock Techniques For The Synthesis Of Multiprocessor Tasksets.
\newblock \emph{WATERS'10}.

\bibitem[{Glover(1989)}]{glover1989tabu}
Glover, F.~W. 1989.
\newblock Tabu Search - Part {I}.
\newblock \emph{{INFORMS} J. Comput.}, 1(3): 190--206.

\bibitem[{Grover et~al.(2019)Grover, Wang, Zweig, and
  Ermon}]{grover2019stochastic}
Grover, A.; Wang, E.; Zweig, A.; and Ermon, S. 2019.
\newblock Stochastic Optimization of Sorting Networks via Continuous
  Relaxations.
\newblock In \emph{Proceedings of the 7th International Conference on Learning
  Representations (ICLR)}.

\bibitem[{Gu and Hao(2018)}]{gupointerkp}
Gu, S.; and Hao, T. 2018.
\newblock A pointer network based deep learning algorithm for 0-1 Knapsack
  Problem.
\newblock In \emph{Proceedings of the 10th International Conference on Advanced
  Computational Intelligence (ICACI)}, 473--477.

\bibitem[{Guan et~al.(2009)Guan, Stigge, Yi, and Yu}]{Guan2009}
Guan, N.; Stigge, M.; Yi, W.; and Yu, G. 2009.
\newblock New Response Time Bounds for Fixed Priority Multiprocessor
  Scheduling.
\newblock In \emph{Proceedings of the 30th {IEEE} Real-Time Systems Symposium
  (RTSS)}, 387--397.

\bibitem[{Gujarati, Cerqueira, and
  Brandenburg(2015)}]{gujarati2015multiprocessor}
Gujarati, A.; Cerqueira, F.; and Brandenburg, B.~B. 2015.
\newblock Multiprocessor real-time scheduling with arbitrary processor
  affinities: from practice to theory.
\newblock \emph{Real Time Syst.}, 51(4): 440--483.

\bibitem[{Gumbel(1954)}]{gumbel1954maxima}
Gumbel, E. 1954.
\newblock The maxima of the mean largest value and of the range.
\newblock \emph{The Annals of Mathematical Statistics}, 76--84.

\bibitem[{Gupta et~al.(2016)Gupta, Liang, Tseng, Vijay, Chen, and
  Rosales}]{emailvolumeoptimize}
Gupta, R.; Liang, G.; Tseng, H.; Vijay, R. K.~H.; Chen, X.; and Rosales, R.
  2016.
\newblock Email Volume Optimization at LinkedIn.
\newblock In \emph{Proceedings of the 22nd {ACM} {SIGKDD} International
  Conference on Knowledge Discovery and Data Mining}, 97--106.

\bibitem[{Hinton, Vinyals, and Dean(2015)}]{hinton2015distilling}
Hinton, G.~E.; Vinyals, O.; and Dean, J. 2015.
\newblock Distilling the Knowledge in a Neural Network.
\newblock \emph{ArXiv}, abs/1503.02531.

\bibitem[{Jang, Gu, and Poole(2017)}]{jang2016categorical}
Jang, E.; Gu, S.; and Poole, B. 2017.
\newblock Categorical Reparameterization with Gumbel-Softmax.
\newblock In \emph{Proceedings of the 5th International Conference on Learning
  Representations (ICLR)}.

\bibitem[{Jiao et~al.(2020)Jiao, Yin, Shang, Jiang, Chen, Li, Wang, and
  Liu}]{jiao-etal-2020-tinybert}
Jiao, X.; Yin, Y.; Shang, L.; Jiang, X.; Chen, X.; Li, L.; Wang, F.; and Liu,
  Q. 2020.
\newblock TinyBERT: Distilling {BERT} for Natural Language Understanding.
\newblock In \emph{Proceedings of the Association for Computational Linguistics
  (EMNLP)}, 4163--4174.

\bibitem[{Kim and Rush(2016)}]{kim2016sequence}
Kim, Y.; and Rush, A.~M. 2016.
\newblock Sequence-Level Knowledge Distillation.
\newblock In \emph{Proceedings of the 2016 Conference on Empirical Methods in
  Natural Language Processing (EMNLP)}, 1317--1327.

\bibitem[{Kool, van Hoof, and Welling(2019)}]{kool2018attention}
Kool, W.; van Hoof, H.; and Welling, M. 2019.
\newblock Attention, Learn to Solve Routing Problems!
\newblock In \emph{Proceedings of the 7th International Conference on Learning
  Representations, (ICLR)}.

\bibitem[{Kool, van Hoof, and Welling(2020)}]{kool2019estimating}
Kool, W.; van Hoof, H.; and Welling, M. 2020.
\newblock Estimating Gradients for Discrete Random Variables by Sampling
  without Replacement.
\newblock In \emph{Proceedings of the 8th International Conference on Learning
  Representations (ICLR)}.

\bibitem[{Lee et~al.(2020)Lee, Lee, Yeom, and Woo}]{lee2020panda}
Lee, H.; Lee, J.; Yeom, I.; and Woo, H. 2020.
\newblock Panda: Reinforcement Learning-Based Priority Assignment for
  Multi-Processor Real-Time Scheduling.
\newblock \emph{{IEEE} Access}, 8: 185570--185583.

\bibitem[{Maddison, Tarlow, and Minka(2014)}]{maddison2014sampling}
Maddison, C.~J.; Tarlow, D.; and Minka, T. 2014.
\newblock A* sampling.
\newblock \emph{arXiv preprint arXiv:1411.0030}.

\bibitem[{Mao et~al.(2016)Mao, Alizadeh, Menache, and
  Kandula}]{mao2016resource}
Mao, H.; Alizadeh, M.; Menache, I.; and Kandula, S. 2016.
\newblock Resource Management with Deep Reinforcement Learning.
\newblock In \emph{Proceedings of the 15th {ACM} Workshop on Hot Topics in
  Networks}, 50--56.

\bibitem[{Mao et~al.(2019)Mao, Schwarzkopf, Venkatakrishnan, Meng, and
  Alizadeh}]{mao2018learning}
Mao, H.; Schwarzkopf, M.; Venkatakrishnan, S.~B.; Meng, Z.; and Alizadeh, M.
  2019.
\newblock Learning scheduling algorithms for data processing clusters.
\newblock In \emph{Proceedings of the {ACM} Special Interest Group on Data
  Communication}, 270--288.

\bibitem[{Mirhoseini et~al.(2017)Mirhoseini, Pham, Le, Steiner, Larsen, Zhou,
  Kumar, Norouzi, Bengio, and Dean}]{mirhoseini2017device}
Mirhoseini, A.; Pham, H.; Le, Q.~V.; Steiner, B.; Larsen, R.; Zhou, Y.; Kumar,
  N.; Norouzi, M.; Bengio, S.; and Dean, J. 2017.
\newblock Device Placement Optimization with Reinforcement Learning.
\newblock In \emph{Proceedings of the 34th International Conference on Machine
  Learning (ICML)}, volume~70, 2430--2439.

\bibitem[{Perron and Furnon(2019-7-19)}]{ortools}
Perron, L.; and Furnon, V. 2019-7-19.
\newblock OR-Tools.

\bibitem[{Reddi et~al.(2021)Reddi, Pasumarthi, Menon, Rawat, Yu, Kim, Veit, and
  Kumar}]{Reddi2021RankDistilKD}
Reddi, S.~J.; Pasumarthi, R.~K.; Menon, A.~K.; Rawat, A.~S.; Yu, F.~X.; Kim,
  S.; Veit, A.; and Kumar, S. 2021.
\newblock RankDistil: Knowledge Distillation for Ranking.
\newblock In \emph{Proceedings of the 24th International Conference on
  Artificial Intelligence and Statistics (AISTATS)}, volume 130, 2368--2376.

\bibitem[{Rendle et~al.(2009)Rendle, Freudenthaler, Gantner, and
  Schmidt{-}Thieme}]{rendle2012}
Rendle, S.; Freudenthaler, C.; Gantner, Z.; and Schmidt{-}Thieme, L. 2009.
\newblock {BPR:} Bayesian Personalized Ranking from Implicit Feedback.
\newblock In \emph{Proceedings of the 25th Conference on Uncertainty in
  Artificial Intelligence}, 452--461.

\bibitem[{Shojaei et~al.(2009)Shojaei, Ghamarian, Basten, Geilen, Stuijk, and
  Hoes}]{shojaei2009parameterized}
Shojaei, H.; Ghamarian, A.; Basten, T.; Geilen, M.; Stuijk, S.; and Hoes, R.
  2009.
\newblock A parameterized compositional multi-dimensional multiple-choice
  knapsack heuristic for CMP run-time management.
\newblock In \emph{Proceedings of the 46th Annual Design Automation
  Conference}, 917--922.

\bibitem[{Silver et~al.(2017)Silver, Schrittwieser, Simonyan, Antonoglou,
  Huang, Guez, Hubert, Baker, Lai, Bolton et~al.}]{silver2017mastering}
Silver, D.; Schrittwieser, J.; Simonyan, K.; Antonoglou, I.; Huang, A.; Guez,
  A.; Hubert, T.; Baker, L.; Lai, M.; Bolton, A.; et~al. 2017.
\newblock Mastering the game of go without human knowledge.
\newblock \emph{nature}, 550(7676): 354--359.

\bibitem[{Tang and Wang(2018)}]{tangrankdistil}
Tang, J.; and Wang, K. 2018.
\newblock Ranking Distillation: Learning Compact Ranking Models With High
  Performance for Recommender System.
\newblock In \emph{Proceedings of the 24th {ACM} {SIGKDD} International
  Conference on Knowledge Discovery {\&} Data Mining (KDD)}, 2289--2298.

\bibitem[{Touvron et~al.(2021)Touvron, Cord, Douze, Massa, Sablayrolles, and
  Jegou}]{pmlr-v139-touvron21a}
Touvron, H.; Cord, M.; Douze, M.; Massa, F.; Sablayrolles, A.; and Jegou, H.
  2021.
\newblock Training data-efficient image transformers \& distillation through
  attention.
\newblock In \emph{Proceedings of the 38th International Conference on Machine
  Learning (ICML)}, volume 139, 10347--10357.

\bibitem[{Vaswani et~al.(2017)Vaswani, Shazeer, Parmar, Uszkoreit, Jones,
  Gomez, Kaiser, and Polosukhin}]{vaswani2017attention}
Vaswani, A.; Shazeer, N.; Parmar, N.; Uszkoreit, J.; Jones, L.; Gomez, A.~N.;
  Kaiser, L.; and Polosukhin, I. 2017.
\newblock Attention is All you Need.
\newblock In \emph{Proceedings of the advances in Neural Information Processing
  Systems (NeurIPS)}, 5998--6008.

\bibitem[{Vinyals et~al.(2019)Vinyals, Babuschkin, Czarnecki, Mathieu, Dudzik,
  Chung, Choi, Powell, Ewalds, Georgiev, Oh, Horgan, Kroiss, Danihelka, Huang,
  Sifre, Cai, Agapiou, Jaderberg, Vezhnevets, Leblond, Pohlen, Dalibard,
  Budden, Sulsky, Molloy, Paine, G{\"{u}}l{\c{c}}ehre, Wang, Pfaff, Wu, Ring,
  Yogatama, W{\"{u}}nsch, McKinney, Smith, Schaul, Lillicrap, Kavukcuoglu,
  Hassabis, Apps, and Silver}]{vinyals2019grandmaster}
Vinyals, O.; Babuschkin, I.; Czarnecki, W.~M.; Mathieu, M.; Dudzik, A.; Chung,
  J.; Choi, D.~H.; Powell, R.; Ewalds, T.; Georgiev, P.; Oh, J.; Horgan, D.;
  Kroiss, M.; Danihelka, I.; Huang, A.; Sifre, L.; Cai, T.; Agapiou, J.~P.;
  Jaderberg, M.; Vezhnevets, A.~S.; Leblond, R.; Pohlen, T.; Dalibard, V.;
  Budden, D.; Sulsky, Y.; Molloy, J.; Paine, T.~L.; G{\"{u}}l{\c{c}}ehre,
  {\c{C}}.; Wang, Z.; Pfaff, T.; Wu, Y.; Ring, R.; Yogatama, D.; W{\"{u}}nsch,
  D.; McKinney, K.; Smith, O.; Schaul, T.; Lillicrap, T.~P.; Kavukcuoglu, K.;
  Hassabis, D.; Apps, C.; and Silver, D. 2019.
\newblock Grandmaster level in StarCraft {II} using multi-agent reinforcement
  learning.
\newblock \emph{Nat.}, 575(7782): 350--354.

\bibitem[{Williams(1992)}]{williams1992simple}
Williams, R.~J. 1992.
\newblock Simple Statistical Gradient-Following Algorithms for Connectionist
  Reinforcement Learning.
\newblock \emph{Mach. Learn.}, 8: 229--256.

\bibitem[{Ykman{-}Couvreur et~al.(2006)Ykman{-}Couvreur, Nollet, Catthoor, and
  Corporaal}]{ykman2006fast}
Ykman{-}Couvreur, C.; Nollet, V.; Catthoor, F.; and Corporaal, H. 2006.
\newblock Fast Multi-Dimension Multi-Choice Knapsack Heuristic for MP-SoC
  Run-Time Management.
\newblock In \emph{Proceedings of the 4th international Symposium on
  System-on-Chip (SoC)}, 1--4.

\bibitem[{Zhang and Dietterich(1995)}]{zhangjobshop}
Zhang, W.; and Dietterich, T.~G. 1995.
\newblock A Reinforcement Learning Approach to job-shop Scheduling.
\newblock In \emph{Proceedings of the 14th International Joint Conference on
  Artificial Intelligence (IJCAI)}, 1114--1120.

\bibitem[{Zoph and Le(2017)}]{zoph2016neural}
Zoph, B.; and Le, Q.~V. 2017.
\newblock Neural Architecture Search with Reinforcement Learning.
\newblock In \emph{Proceedings of the 5th International Conference on Learning
  Representations (ICLR)}.

\end{thebibliography}


\appendix
\renewcommand{\theequation}{A.\arabic{equation}}
\renewcommand{\thefigure}{A.\arabic{figure}}
\renewcommand{\thetable}{A.\arabic{table}}
\setcounter{secnumdepth}{1}
\setcounter{equation}{0}
\section{Proof of proposition 1} \label{apsec:proof}
\begin{proof}
By the chain rule, we have 
\begin{equation*}
 \frac{\partial g}{\partial \textbf{x}}(\textbf{s}) = \frac{\partial{L}}{\partial{\textbf{y}}}(\tilde{\textbf{rk}}_\epsilon(\textbf{s})) \frac{\partial{\tilde{\textbf{rk}_{\epsilon}}}}{\partial{\textbf{x}}}(\textbf{s})
\end{equation*}
where the first part $\frac{\partial{L}}{\partial{\textbf{y}}}(\tilde{\textbf{rk}}_\epsilon(\textbf{s})) = [e_1, \cdots, e_n]$ is followed directly from the elementary calculus. 
For the second part, we note that $\tilde{\textbf{rk}}_{\epsilon} = P(-\frac{\textbf{s}}{\epsilon}, \tau)$ where $\tau = [n, \cdots, 1]$ and $P(\textbf{z}, \textbf{w}) = \textnormal{ArgMin}\{\frac{1}{2} \lVert \textbf{y} - \textbf{z} \rVert \mid \textbf{y} \in \mathcal{P}(\textbf{w})\}$  
where $\mathcal{P}(\textbf{w})$ is a permutahedron generated by $\textbf{w}$.
Applying the chain rule and proposition 4 in~\cite{blondel2020fast}, we obtain 
\begin{equation*}
    \frac{\partial \tilde{\textbf{rk}}_\epsilon}{\partial \textbf{x}}(\textbf{s}) = -\frac{\textbf{I}}{\epsilon}\left( \textbf{I} - 
	\begin{bmatrix}
		\frac{1}{n_1}\textbf{1}_{n_1} & \textbf{0} & \textbf{0} \\
		\textbf{0} & \dots & \textbf{0} \\
		\textbf{0} & \textbf{0} & \frac{1}{n_m}\textbf{1}_{n_m}
	\end{bmatrix}
	  \right)_{\phi}
\end{equation*}
for some integers $n_1, \cdots, n_m$ such that $n_1 + \cdots + n_m = n$ and 
some permutation $\phi$. This completes the proof.
\end{proof}

Here, we can analyze how $g$ behaves as we tune the hyperparameter $\epsilon$. We first formulate the function $\tilde{\textbf{rk}}_\epsilon$ in terms of an isotonic optimization problem,  
\begin{align}
	\textbf{v}(-\frac{\textbf{s}}{\epsilon}, \tau) &= 
	\underset{v_1 \geq \cdots \geq v_n}{\textnormal{ArgMin}} \lVert \textbf{v} - (\textbf{s} - \tau) \rVert ^2
	~\label{argmin} \\
	& = \underset{v_1 \geq \cdots \geq v_n}{\textnormal{ArgMin}} \sum_{j=1}^n (v_j - (-\frac{s_j}{\epsilon} - (n-j+1)))^2
	\label{argmin2}
\end{align}
where $\textbf{v} = (v_1, \cdots, v_n)$ and $\textbf{s} = (s_1, \cdots, s_n)$. It is known that $\tilde{\textbf{rk}}_\epsilon(\textbf{s}) = -\frac{\textbf{s}}{\epsilon} - \textbf{v}(-(\frac{\textbf{s}}{\epsilon})_\phi, \tau)_{\phi^{-1}}$ for some $\phi$. 
Note that the optimal solution of $p$th term inside the ArgMin of~\eqref{argmin2} is just $o_p = -\frac{s_p}{\epsilon} - (n-p+1)$. 
Fix a vector $\textbf{s}$ for a moment in the following. 

If we take $\epsilon \longrightarrow \infty$, then the values $o_p$ become more out-of-order, meaning that $o_p < o_{p+1}$. This implies the partition $\mathcal{B}_1, \cdots, \mathcal{B}_m$ of $\{1, \cdots, n\}$ appearing in a pool adjacent violator (PAV) algorithm~\cite{pavalgorithm} becomes more chunky (Note that our objective in ~\eqref{argmin} is ordered in decreasing). This implies the block diagonal matrix in the right hand side of (20) in Proposition 1 becomes more uniform. When the right hand side form is multiplied with the reciprocal of $\epsilon$ in the same equation, the gradient $\frac{\partial{g}}{\partial \textbf{s}}$ is the small enough, but in the uniform manner.

On the other hand, if we take $\epsilon \longrightarrow 0$, the subtraction of block diagonal matrix from $\textbf{I}$ has more zero entries. This is because the optimization$~\eqref{argmin}$ has much more chance to have in-order solutions, meaning that $o_p \geq o_{p+1}$. Consequently, the block diagonal matrix in (20) tends to be an identity matrix. In the training phase, the error term $e_i$ becomes small enough if the score $\textbf{s}$ leads to correct predictions for given ranking labels. Accordingly,  the $i$th score is not engaged in the backpropagation. 

In summary, we exploit $\epsilon$-controllability such that ``hard'' rankings with exact gradient and  ``soft'' rankings with uniform, small gradient for different $\epsilon$ settings from $0$ to $\infty$.

\renewcommand{\theequation}{B.\arabic{equation}}
\renewcommand{\thefigure}{B.\arabic{figure}}
\renewcommand{\thetable}{B.\arabic{table}}

\setcounter{equation}{0}
\section{Multidimensional Knapsack Problem (MDKP)} \label{apsec:knapsack}

\subsection{Problem Description}
%
%
%
%



Given an $N$-sized item set $\mathcal{I}$ where the $j$th item is represented by $x_j = [v^{(j)}, w^{(j)}_1, \cdots, w^{(j)}_k] \in \mathbb{R}^{k+1}$, $x_j^{(v)} = v^{(j)}$ is a value and $x_j^{(w)} = [w_1^{(j)}, \cdots, w_k^{(j)}]$ represent $k$-dimensional weights. A knapsack has $k$-dimensional capacities $C = [c_1, \cdots, c_k]$. 

In RL formulation, a state consists of $O_t$ and $S_t$ at timestep $t$ such that $\mathcal{I} = O_t \cup S_t$. Note that $O_t$ is a set of allocable items and $S_t$ is a set of the other items in $\mathcal{I}$. 
To define an allocable item, suppose the items $x_{i_1}, \cdots, x_{i_{t-1}}$ are in the knapsack at time $t$. Then, an item $x_j $ is allocable if and only if $j \notin \{i_1, \cdots, i_{t-1}\}$ and $x_j^{(w)} \leq C - \sum_{j=1}^{t-1}x_{i_j}^{(w)}$. Then, an action $a_t = j$ corresponds to a selection to place $x_j$ in the knapsack. Considering the objective of MKDP maximizing the total value of selected items, a reward is given by 
\begin{align}
	r_t =
	\begin{cases}
		0 & \textnormal{if} \,\,  \exists \,\textnormal{allocable items} \\
		\sum_{i \in \textnormal{sack}} x_i^{{(v})}	&	\textnormal{otherwise.}
	\end{cases}
\end{align}
When no items remain to be selected, an episode terminates.
Accordingly, a model is learned to establish policy $\pi_\theta$  that maximizes the total reward in (11) and thus optimizes the performance of total values achieved for a testing set, e.g.,  
\begin{align}
	\sum_{\mathcal{I} \in \mathcal{D}} \sum_{t=1}^E x_{a_t}^{(v)}
\end{align}
where $\mathcal{D}$ is a testing set of item sets, $a_t = j$ represents the index of a selected item, and $E$ denotes a length of an episode. We set each episode to have an single item set with $N$ items.



\subsection{Dataset Generation}
Given a number of items $N$, we set $k$-dimensional weights $x_j^{(w)}$ of item to be randomly generated from a uniform distribution $\uniform(1, w)^k$ for integers, 
and, the item value is given by 
\begin{equation}
	x_j^{(v)} = \alpha*\frac{\sum _{j=1}^k x_j^{(w)}} {k} + (1 - \alpha)*\textbf{Unif}(1, 200)
\end{equation}
where $\alpha \in \{0, 0.9\}$ manipulates the correlation of weights and values.
For $k$-dimensional capacities $C$ of a given knapsack, we set 
$C = 0.5 * \sum_{i\in \mathcal{I}} x_i^{(w)}$. A range of parameter values for items is given in Table~\ref{tbl:mdkpdata}.

\begin{table}[h]
    \centering
    \begin{tabular}{|c||c|}
    \hline
    Item Parameter  &  Values \\ \hline
    Num. of items $N$     &  \{50, 100, 150\} \\ \hline
    Weights dim. $k$     & \{3, 10, 15\} \\ \hline
    Max weights $w$     & \{200, 500\} \\ \hline
    Correlation $\alpha$ & \{0, 0.9\} \\ \hline
    \end{tabular}
    \caption{The generation range for items in MDKP}
    \label{tbl:mdkpdata}
\end{table}

Through the encoder, input items are transformed to a raw feature vector representation. Given vector $\textbf{z} = [z_1, \cdots, z_n]$, we denote \textbf{Max(\textbf{z})} = max$\{z_1, \cdots, z_n\}$, \textbf{Min}(\textbf{z}) = min$\{z_1, \cdots, z_n\}$, and \textbf{Mean(\textbf{z})} = $\frac{1}{n} \sum_i z_i$. For item $x_i$, we denote its utilization as $\textbf{Ut}(x_i)= \frac{x_i^{(w)}}{C}$. These input raw features are listed in Table~\ref{input_tr}.

\begin{table}[h]

	\centering
	\begin{adjustbox}{width=0.47\textwidth}
	\begin{tabular}{|c||c|}
	\hline
	Item representation & $x_i^{(w)}$ \\ \hline
	Weight Utilization & $\textbf{Mean}(\textbf{Ut}(x_i))$, $\textbf{Max}(\textbf{Ut}(x_i))$, $\textbf{Min}(\textbf{Ut}(x_i))$ \\ \hline
	Value Utilization	&  $x_i^{(v)} / \textbf{Mean}(\textbf{Ut}(x_i))$, \,\, $x_i^{(v)} / \textbf{Max}(\textbf{Ut}(x_i))$, \,\, $x_i^{(v)} / \textbf{Min}(\textbf{Ut}(x_i))$
	\\ \hline
	Min-Max Utilization & $\frac{\textbf{Mean}(\textbf{Ut}(x_i))}{\textbf{Max}(\textbf{Ut}(x_i))}$, $\frac{\textbf{Mean}(\textbf{Ut}(x_i))}{\textbf{Min}(\textbf{Ut}(x_i))}$, $\frac{\textbf{Max}(\textbf{Ut}(x_i))}{\textbf{Min}(\textbf{Ut}(x_i))}$ \\ \hline
	\end{tabular}
	\end{adjustbox}
	\caption{Raw features for items in MDKP}
	\label{input_tr}
\end{table}

\subsection{Model Structure}

In MDKP, we use an encoder structure for a student network in the RL-to-rank distillation. Specifically, item $x_i$ is converted to the embedding $h_i = Wx_i + b$ for some parameters $W$ and $b$, giving a matrix $H^{(0)} = [h_1, \cdots, h_N]$. 
A global representation on the embeddings is calculated by 
\begin{align}
    x^{(g)} = l^{(0)} \odot \textbf{Att}(l^{(0)}, H^{(0)})
\end{align}
for arbitrarily initialized learnable parameter $l^{(0)}$.
Then, the score of item $x_i$ is defined by
\begin{align}
    \textbf{SCORE}(x_i) = 10 * (\textbf{tanh} (\textbf{Att}(x^{(g)}), H^{(0)}))_i    
\end{align}
which is similar to (6) for timestep $t=1$. The detail hyper parameter settings for the teacher RL and distilled student models are given in Table~\ref{hyperparameters_MDKP}

\begin{table}[h]
    \centering
    \begin{adjustbox}{width=0.3\textwidth}
    \begin{tabular}{|c||c|c|}
    \hline
     & \multicolumn{1}{c|}{RL } & Distilled \\ \hline\hline

	Batch size & 128 & 1 \\ \cline{1-3}
	Iterations & 250 & 10000 \\ \cline{1-3}
	Learning Rate & $5 * 10^{-3}$ & $5 * 10^{-3}$ \\ \cline{1-3}
	Optimizer & Adam & Adam \\ \cline{1-3}
	$\epsilon$ of DiffRank & - &$10^{-3}$ \\ \cline{1-3}
	Embed. Dim. & 256 & - \\ \cline{1-3}
	Num. Att. Layer & 2 & - \\ \cline{1-3}
	Discount Factor $\gamma$ & 1 & - \\ \hline
    \end{tabular}
    \end{adjustbox}
    \caption{Hyperparameter settings in MDKP}
    \label{hyperparameters_MDKP}
\end{table}

\subsection{Model Training}

In training with the REINFORCE algorithm, we leverage greedy algorithms to establish a baseline $b(t)$ in (12). Suppose that by using a specific greedy algorithm that selects an item of the highest weight-value ratio at each timestep, say greedy policy $\psi$, we obtain an episode $T^{(\psi)} = (s_1^{(\psi)}, a_1^{(\psi)}, r_1^{(\psi)} \cdots, s_E^{(\psi)}, a_E^{(\psi)}, r_E^{(\psi)})$. Then, the baseline $b(t)$ is established upon the episode by 
\begin{equation}
    b(t) = \sum_{k=t}^E \gamma^{k-t}r_{k}^{(\psi)}.
\end{equation}
for discount factor $\gamma$.

For the problem-specific loss function $\mathcal{L}_P$ in (17), consider a student network generating score vectors $\textbf{s} = [s_1, \cdots, s_N]$. 
Then, if each item is selected by the rankings based on scores in $\textbf{s}$, we obtain a vector $M(\textbf{s}) = [\delta^{(S)}_1, \cdots, \delta^{(S)}_N]$ with
\begin{align}
    \delta_i^{(S)} = 
    \begin{cases}
        1   &   \textnormal{if} \,\, x_i \in \textnormal{knapsack}   \\ 
        0   &   \textnormal{otherwise}.
    \end{cases}
\end{align}

Similarly, if each item is selected by the rankings based on supervised labels $\textbf{y}$ from the teacher RL model, 
we obtain $M(\textbf{y}) = [\delta_1^{(RL)}, \cdots, \delta_N^{(RL)}]$ where $\delta_i^{(RL)} = 1$.   
Accordingly, we can have $\mathcal{L}_P$ such that
\begin{align}
    \mathcal{L}_P(\textbf{y}, \textbf{s}) = \left( M(\textbf{s}) - M(\textbf{y}) \right) \cdot \textbf{s}
\end{align}
that penalizes the difference on $x_i \in \textnormal{knapsack}$  made by the rankings based on the supervised labels and the rankings based on the scores.


We implement our models using Python3 v3.6 and Pytorch v1.8, and train the models on a system of Intel(R) Core(TM) i9-10940X processor, with an NVIDIA RTX 3090 GPU.

\subsection{Comparison with an Optimal Solver}
\begin{table}[h]
    \centering
    \begin{adjustbox}{width=0.47\textwidth}
    \begin{tabular}{|c|cc|cc|cc|cc|}
        \hline
        \multirow{2}{*}{limit} & \multicolumn{2}{c|}{SCIP} & \multicolumn{2}{c|}{GLOP} & \multicolumn{2}{c|}{RD} & 
        \multicolumn{2}{c|}{RD-G} \cr
        & - & Time & Perf & Time & Perf & Time & Perf & Time \cr \hline 
        3s & 100\% & 3s & 894\% & 0.02s & 884\% & 0.004s & 908\% & 0.02s \cr
        60s & 100\% & 60s & 201\% & 0.02s & 198\% & 0.004s & 209\% & 0.02s \cr 
        600s & 100\% & 600s & 90.4\% & 0.02s & 92.0\% & 0.004s & 94.7\% & 0.02s \cr \hline
    \end{tabular}
    \end{adjustbox}
    \caption{Relative performance to SCIP}
\label{tbl:sciptab}
\end{table}

We test an optimal mixed integer programming solver SCIP implemented in the OR-tools, and we observe that it has slow inferences as shown in Table~\ref{tbl:sciptab}.
%
For example, SCIP achieves 8 times lower performance than ours (RD, RD-G) under 3s time limit, but it achieves only 5\% higher performance than ours under 600s time limit. As such, SCIP shows insignificant performance gain compared to the unacceptably slow inference time for mission critical environments.



\renewcommand{\theequation}{C.\arabic{equation}}
\renewcommand{\thefigure}{C.\arabic{figure}}
\renewcommand{\thetable}{C.\arabic{table}}
\setcounter{equation}{0}
\section{Global Fixed Priority Scheduling \\ (GFPS)}\label{apsec:GFPS}
\subsection{Problem Description} 
For a set of $N$-periodic tasks, in GFPS, each task is assigned a priority (an integer from $1$ to $N$) to be scheduled. That is, GFPS with a priority order (or a ranking list) is able to schedule the $m$ highest-priority tasks in each time slot upon a platform comprised of $m$ homogeneous processors without incurring any deadline violation of the periodic tasks over time. 


Specifically, given an $N$-sized task set $\mathcal{I}$ where each task is specified by its period, deadline, and worst case execution time (WCET), we formulate an RL procedure for GFPS as below.   
A state is set to have two task sets $O_t$ and $L_t$ for each timestep $t$, where the former specifies priority-assigned tasks and the latter specifies the other tasks in $\mathcal{I}$.  Accordingly, $O_t \cup L_t = \mathcal{I}$ holds for all $t \in \{1,2,\dots,N\}$. 
An action $a_t = i$ ($i \in \{1, 2, \dots, N\}$) implies that the task $x_i$ is assigned the highest priority in $L_t$.
A reward is calculated by 
\begin{equation}
	r_t = 
	\begin{cases}
		0 & \textrm{if}\,\, t \neq N \\
		\textbf{Test}(\mathcal{I}, \phi) & \textrm{otherwise}.
	\end{cases}
\end{equation} 
where $\phi$ is a priority order and $\textbf{Test}(\mathcal{I}, \phi)$ is a given schedulability test~\cite{Guan2009} which maps $\phi$ to 0 or 1 depending on whether or not a task set $\mathcal{I}$ is schedulable by GFPS with $\phi$. 

With those RL components above, a model is learned to establish such a policy $\pi_{\theta}((O_t, L_t), a_t=i)$ in (9) that generates a schedulable priority order for each $\mathcal{I}$, which passes $\textbf{Test}(\mathcal{I}, \phi)$, if that exists. The overall performance is evaluated upon a set of task set samples  $\mathcal{D}$.
\begin{equation} 
   \text{Schedulability Ratio} = \sum_{\mathcal{I} \in \mathcal{D}} \test(\mathcal{I}, \phi) \  / \ |\mathcal{D}|  
   \label{equ:sched_ratio}
\end{equation}
Note that we focus on an implicit deadline problem, meaning that period and deadline are the same, under non-preemptive situations.

\subsection{Dataset Generation}
To generate task set samples, we exploit the~\textit{Randfixedsum} algorithm~\cite{Emberson2010}, which has been used widely in research of scheduling problems~\cite{brandenburg2016global, gujarati2015multiprocessor}. For each task set, we first configure the number of tasks $N$ and total task set utilization $u$. The \textit{Randfixedsum} algorithm randomly selects utilization of $U_i$ for each task $\mathcal{I}$, i.e., $u = \sum_{i=1}^{N} U_i$. The algorithm also generates a set of task parameter samples each of which follows the rules in Table~{\ref{tbl:taskparams}}, yielding values for period $T_i$, WCET $C_i$, and deadline $D_i$ under the predetermined utilization $U_i$, where we have  $U_i = \frac{C_i}{T_i}$. 
%
\begin{table}[h] 
\centering
\begin{tabular}{|c||c|}
    \hline
    Task Parameter & Generation Rule \\ 
    \hline 
    $T_i$ & $\floor{ \exp(\uniform(\log T_{\text{min}},\ \log T_{\text{max}})) }$ \\ \hline
    $C_i$ &   $\floor{T_i U_i}$  \\ \hline
    $D_i$ & $T_i = D_i$ \ \text{for implicit deadline}\\ \hline
\end{tabular}
\caption{Generation rules for tasks in GFPS}
\label{tbl:taskparams}
\end{table}

Three properties $T_i, C_i$ and $D_i$ are transformed into a vector representation of each task in Table~\ref{tbl:taskrep}.
\begin{table}[h]
    \centering
    \begin{adjustbox}{width=0.47\textwidth}
    \begin{tabular}{|c||c|}
    
    \hline
    Task representation     & $T_i, C_i, D_i$ \\ \hline
    Log scale representation     & $\textnormal{log}(1+C_i), \textnormal{log}(1+T_i), \textnormal{log}(1+D_i)$ \\ \hline
    Utilization & $C_i / T_i, C_i / D_i, D_i / T_i$ \\ \hline
    Slack time & $T_i - C_i, T_i - D_i, D_i - C_i$ \\ \hline
    \end{tabular} 
    \label{tab:my_label}
    \end{adjustbox}
    \caption{Raw features for tasks in GFPS}
    \label{tbl:taskrep}
\end{table}

\subsection{Model structure}

In GFPS, we use a linear network for a student model. Specifically, task $x_i$ is transformed to its corresponding score by
\begin{align}
    \textbf{SCORE}(x_i) = Wx_i + b
\end{align}
for some learnable parameter $W$ and $b$. Hyperparameter settings for RL and distilled models are shown in Table \ref{tbl:hyperparameters}

\begin{table}[h]
    \centering
    \begin{adjustbox}{width=0.35\textwidth}
    \begin{tabular}{|c||c|c|}
    \hline
     & \multicolumn{1}{c|}{RL} & Distilled  \\ \hline \hline
	\#Train. Samples & 200K & 200K \\ \cline{1-3}
	Batch size & 256 & 20 \\ \cline{1-3}
	Epochs & 30 with early stop & 9 \\ \cline{1-3}
	Learning Rate & $10^{-4}$ & $5 * 10^{-3}$ \\ \cline{1-3}
	Optimizer & Adam & Adam \\ \cline{1-3}
	$\epsilon$ of DiffRank & - & $10^{-3}$ \\ \cline{1-3}
    Embedded Dim. & 128 & - \\  \cline{1-3}
    Num. Att. Layer & 2 & - \\ \cline{1-3}
    Discount Factor $\gamma$ & 1 & - \\ \hline
    \end{tabular}
    \end{adjustbox}
    \caption{Hyperparameter settings in GFPS}
    \label{tbl:hyperparameters}
\end{table}

\renewcommand{\theequation}{D.\arabic{equation}}
\renewcommand{\thefigure}{D.\arabic{figure}}
\renewcommand{\thetable}{D.\arabic{table}}
\setcounter{equation}{0}

\section{Travelling Salesman Problem \\(TSP)}\label{apsec:tsp}
TSP is a problem of assigning priorities to $N$ points to be visited so that the total distance to visit all points is minimized.
Specifically, given an $N$-sized set of points $\mathcal{I}$, a state consists of two sets $O_t$ and $L_t$ at each timestep $t$, where $O_t$ is a set of priority-assigned points and $L_t$ is priority-unassigned points. An action $a_t = i$ where $i \in L_t$ corresponds to assign the highest rank to points $x_i$, so that it is visited at time $t+1$. We define a reward by 
\begin{align}
    r_t = 
    \begin{cases}
        0   &   \textnormal{if} \,\, t \neq N \\
        \textbf{-Dist}(\mathcal{I}, \phi)    & \textnormal{otherwise}
    \end{cases}
\end{align}
where $\phi$ is a priority order of all points, and the function \textbf{Dist} calculates the total distance by $\phi$.

Here, we report the TSP performance of RL-based models, compared to a random insertion heuristic method in Table~\ref{tbl:perf_TSP}. For this experiment on TSP, we use the open source implemented in~\cite{kool2018attention}, which has an encoder-decoder structure with sequential processing, similar to ours. 
\begin{table}[h]
    \centering
    \begin{adjustbox}{width=0.47\textwidth}
    \begin{tabular}{|c||ccc|ccc|}
    \hline
     & \multicolumn{3}{c|}{Heuristics} & \multicolumn{3}{c|}{RL} \\
     & Distance & Gap & Time(s) & Distance & Gap & Time(s)
     
     \\ \hline\hline

	N=20 & 3.92 & - &  0.0019 & 3.84 &97.9\% & 0.4704\\ \hline
	N=50 & 6.01 & - &  0.0054 & 5.89 & 98.0\% & 0.9709\\ \hline
	    \end{tabular}
    \end{adjustbox}
    \caption{Performance of TSP. A Gap is a ratio of distance to Heuristics.}
    \label{tbl:perf_TSP}
\end{table}


\section{Limitations and Generality of Our Approach}
We analyze the characteristic of various COPs in Figure~2 in the main manuscript, demonstrating that \textbf{RLRD} works well for MDKP and GFPS, but does not for TSP. Here, we provide more detail explanation by comparing TSP with \textit{Sorting}.
%

Let $c$ be  $[c_1, c_2, c_3] = [4, 1, 2]$ and consider the corresponding unnormalized distribution $p_c = [\exp( c_1), \exp( c_2), \exp( c_3)]$. Upon sampling $4$ which is the number with the largest probability in $p_c$ greedily from $p_c$, a new set $c' = [1, 2]$ is obtained and has to be sorted.
By masking the index of 4 in $p_c$, new unnormalized probability distribution $p_{c'} = [0, \exp( c_2), \exp( c_3)]$ is obtained. Sorting $c'$ is conducted by sampling $2$ greedily from $p_{c'}$. 
%
%
In other words, sorting can be perfectly solved by inductively doing greedy sampling without replacement, starting from one single distribution $p_c$.
On the other hand, in TSP, the probability distribution over items highly depends on the items chosen at previous steps, or contexts, as we described.
In this sense, we consider TSP to be \textit{fully context-dependent}, because no such single distribution $p_c$ exists for TSP in contrast to sorting.
%
If the student model performs greedy sampling without replacement from one distribution $p_c$ for a \textit{context-dependent} problem, it might end up with sub-optimal performance.
%

Many COPs lie in between sorting (\textit{no context-dependent}) and TSP (\textit{fully context-dependent}) in terms of the degree of context dependency, and we consider them as a feasible problem space for RLRD.
We present such characteristics of MDKP and GFPS compared with TSP, which are summarized in the table below.

\begin{table}[H]
\centering
\begin{adjustbox}{width=0.47\textwidth}
\begin{tabular}{c|ccc}
\hline
COP                            & MDKP & GFPS & TSP  \\ \hline
Context-Dependency             & Low  & Moderate   & High \\
Performance of Greedy Baseline & High & Moderate   & Low  \\ \hline
\end{tabular}
\end{adjustbox}
\caption{Characteristics of COPs}
\label{tbl:yourt}
\end{table}

\renewcommand{\theequation}{E.\arabic{equation}}
\renewcommand{\thefigure}{E.\arabic{figure}}
\renewcommand{\thetable}{E.\arabic{table}}

\renewcommand{\theequation}{F.\arabic{equation}}
\renewcommand{\thefigure}{F.\arabic{figure}}
\renewcommand{\thetable}{F.\arabic{table}}

\setcounter{equation}{0}

\section{Choice of the Number of Sampling}

In this section, we explain our default settings on the number of Gumbel samplings, which are set to 30 and 10 for MDKP and GFPS problems,  respectively.
\begin{figure}[h]
\centering
\begin{subfigure}{1.0\linewidth}
    \includegraphics[width=0.9\linewidth]{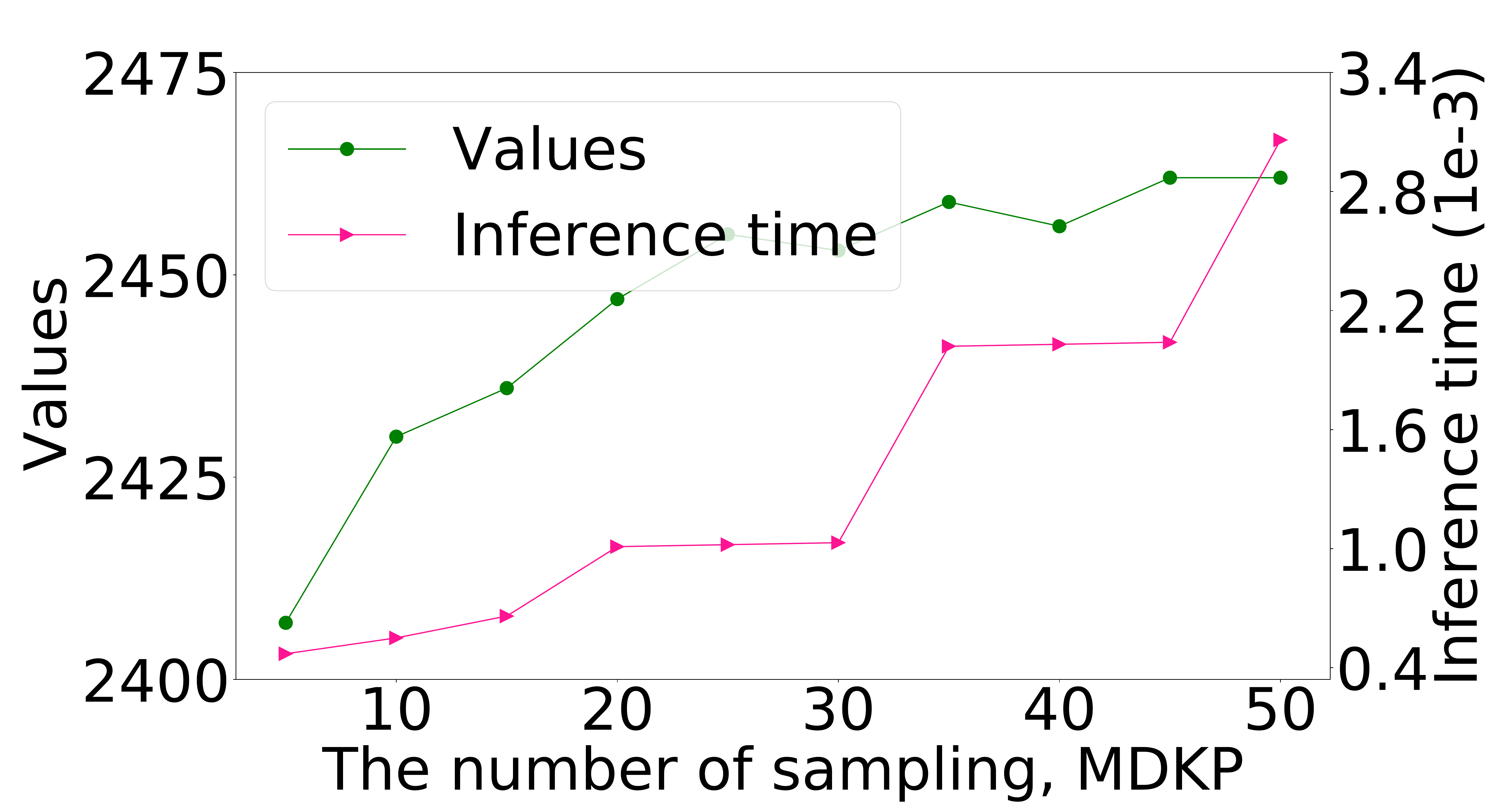}
	\end{subfigure}
\begin{subfigure}{1.0\linewidth}
    \includegraphics[width=0.9\linewidth]{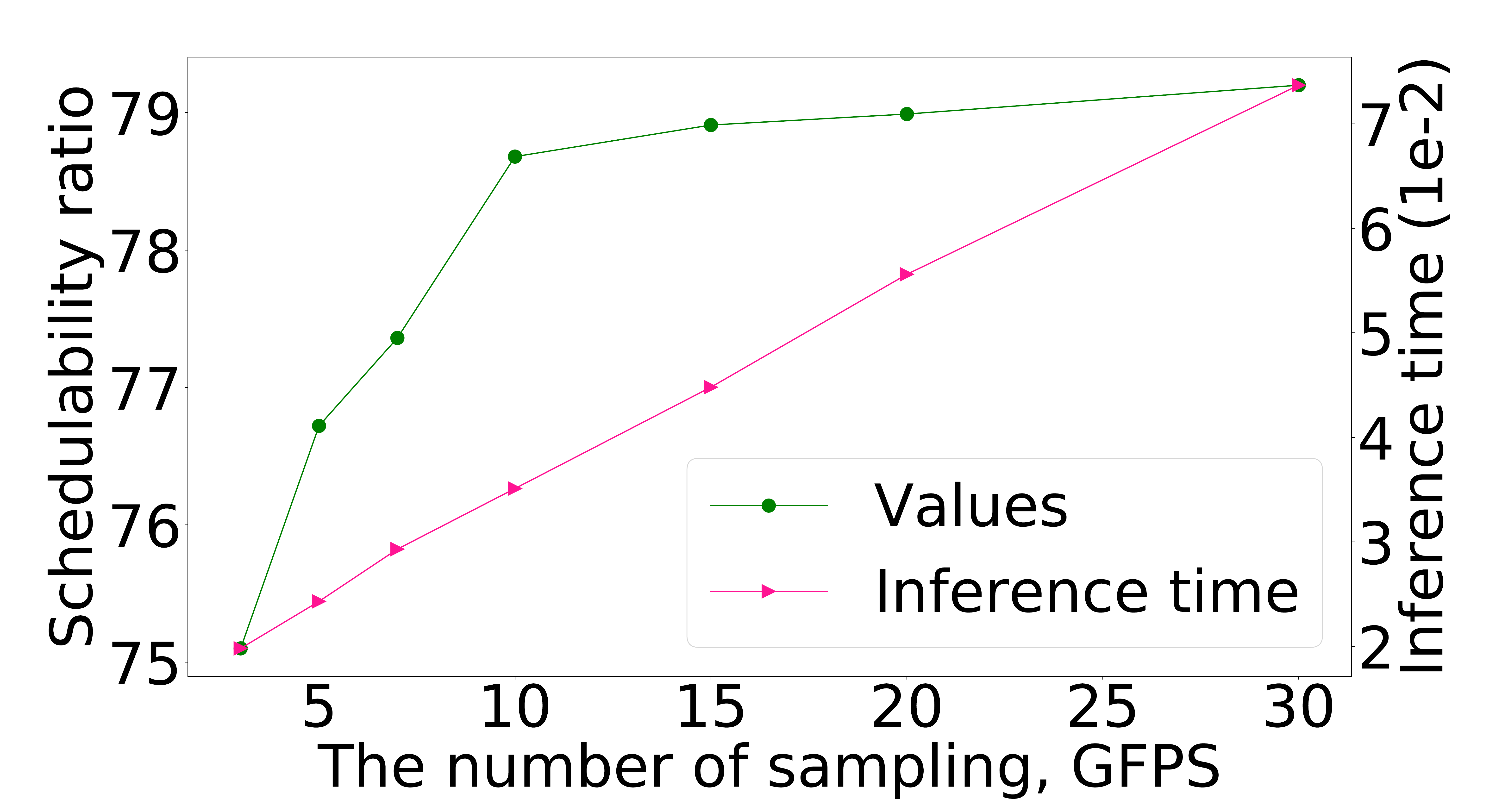}
	\end{subfigure}
	\caption{Overall performance and inference time with respect to Gumbel sampling sizes}
	\label{fig:gumnum}
\end{figure} 

Figure~\ref{fig:gumnum} shows the performance and inference time in MDKP and GFPS, with respect to Gumbel sampling sizes.
In MDKP, the performance gain trading off the inference time decreases around 30 samples. Likewise, in GFPS, the performance gain  trading off the inference time decreases around 10 samples. These empirical results attribute to our settings for  the Gumbel trick-based sequence sampling. We also consider the inference time of teacher RL models, and limit these sampling sizes to make the inference time of distilled models much shorter than their respective teacher RL models.  

Furthermore, we normalize score $\textbf{s} = [s_1, s_2, \cdots, s_N]$ by 
\begin{equation}
    s_i \longleftarrow \frac{N * s_i }{\textbf{max}(\textbf{s}) - \textbf{min}(\textbf{s})} \, \, \, \, \, i \in \{1, \cdots, N\}
\end{equation}
before Gumbel perturbation is added to score $\textbf{s}$, shown in (21). This is because the Gumbel-trick might not generate any variation of rankings, if the gap between item scores is too big, and Gumbel perturbation might generate irrelevant random rankings, if the gap is too small.


\end{document}